\documentclass[paper]{JHEP3}
\pdfoutput=1
\usepackage{amsmath,amssymb,amsthm,amscd,graphicx}
\input epsf.sty

\theoremstyle{definition}

\newcommand{\CA}{{\cal A}}

\newcommand{\CC}{{\cal C}}

\newcommand{\CN}{{\cal N}}
\newcommand{\CO}{{\cal O}}
\newcommand{\CP}{{\cal P}}

\def\IZ{{\mathbb Z}}
\def\IR{{\mathbb R}}
\def\IC{{\mathbb C}}
\def\IP{{\mathbb P}}

\newcommand{\re}{{\rm e}}
\newcommand{\ri}{{\rm i}}
\newcommand{\rd}{{\rm d}}

\newcommand{\dd}{{\rm d}}

\newcommand{\be}{\begin{equation}}
\newcommand{\ee}{\end{equation}}
\newcommand{\ba}{\begin{aligned}}
\newcommand{\ea}{\end{aligned}}
\newcommand{\ben}{\begin{eqnarray}\displaystyle}
\newcommand{\een}{\end{eqnarray}}

\newcommand{\sectiono}[1]{\section{#1}\setcounter{equation}{0}}

\newdimen\tableauside\tableauside=1.0ex
\newdimen\tableaurule\tableaurule=0.4pt
\newdimen\tableaustep
\def\phantomhrule#1{\hbox{\vbox to0pt{\hrule height\tableaurule width#1\vss}}}
\def\phantomvrule#1{\vbox{\hbox to0pt{\vrule width\tableaurule height#1\hss}}}
\def\sqr{\vbox{%
  \phantomhrule\tableaustep
  \hbox{\phantomvrule\tableaustep\kern\tableaustep\phantomvrule\tableaustep}%
  \hbox{\vbox{\phantomhrule\tableauside}\kern-\tableaurule}}}
\def\squares#1{\hbox{\count0=#1\noindent\loop\sqr
  \advance\count0 by-1 \ifnum\count0>0\repeat}}
\def\tableau#1{\vcenter{\offinterlineskip
  \tableaustep=\tableauside\advance\tableaustep by-\tableaurule
  \kern\normallineskip\hbox
    {\kern\normallineskip\vbox
      {\gettableau#1 0 }%
     \kern\normallineskip\kern\tableaurule}%
  \kern\normallineskip\kern\tableaurule}}
\def\gettableau#1{\ifnum#1=0\let\next=\null\else
\squares{#1}\let\next=\gettableau\fi\next}

\newcommand{\figref}[1]{Fig.~\protect\ref{#1}}

\title{Resumming the string perturbation series}

\author{
Alba Grassi, Marcos Mari\~no and Szabolcs Zakany
\\
D\'epartement de Physique Th\'eorique et Section de Math\'ematiques,\\
Universit\'e de Gen\`eve, Gen\`eve, CH-1211 Switzerland\\
\\
\email{alba.grassi@unige.ch, marcos.marino@unige.ch, szabolcs.zakany@unige.ch}
}

\abstract{We use the AdS/CFT correspondence to study the 
resummation of a perturbative genus expansion appearing in the type II superstring dual of ABJM theory. Although the series is Borel summable, 
its Borel resummation does not agree with the exact non-perturbative answer due to the presence of complex instantons. The same type of behavior 
appears in the WKB quantization of the quartic oscillator in Quantum Mechanics, which we analyze in detail as a toy model for the string perturbation series. 
We conclude that, in these examples, Borel summability is not enough for extracting non-perturbative information, due to non-perturbative effects associated to 
complex instantons. We also analyze the resummation of the genus expansion for topological string theory on local $\IP^1 \times \IP^1$, which is closely related to ABJM theory. 
In this case, the non-perturbative answer involves membrane instantons computed by the refined topological string, which are crucial to produce a well-defined 
result. We give evidence that the Borel resummation of the perturbative series requires such a non-perturbative sector.}

\begin{document}

\sectiono{Introduction}

Most of the perturbative series appearing in quantum theory are asymptotic rather than convergent. Therefore, the question arises of how 
to make sense of the information that they encode in order to reconstruct the underlying physical quantities. 
A powerful technique to handle this problem is the theory of Borel transforms and Borel resummation (see \cite{caliceti,mmreview} for reviews). 
In favorable situations, this procedure makes sense of the original perturbative series and leads to a well-defined result, at least for some values of the coupling constant. 
In practice, it can be often combined with the theory of Pad\'e approximants into what we will call the Borel--Pad\'e resummation method. Using this method, 
one can in principle obtain precise numerical values from the asymptotic series, and increase the accuracy 
of the calculation by incorporating more and more 
terms, exactly as one would do with a convergent series. 

The procedure of Borel resummation has been applied successfully in many problems in Quantum Mechanics and in Quantum Field Theory. For example, 
the perturbative series for the energy levels of the quartic anharmonic oscillator is known to be divergent for all values of the coupling \cite{bw1, bw2}, 
yet its Borel resummation can be performed 
and it agrees with the exact values obtained from the Schr\"odinger equation \cite{ggs} (see \cite{caliceti} for a review). 
In this case, the series has the property of being Borel summable, which means roughly that no singularities 
are encountered in the process of Borel resummation. However, in many cases of interest, the divergent series is not Borel summable:  singularities are encountered, and they 
lead to ambiguities in the Borel resummation. These ambiguities are exponentially small and invisible in perturbation theory, and they signal 
the existence of non-perturbative effects. In order to cure these ambiguities, one needs to include instanton sectors 
(or other type of non-perturbative information) to reconstruct the exact answer. The canonical 
example of this situation is the double-well potential in Quantum Mechanics \cite{zj,zjj}, although there are simpler examples in the theory 
of Painlev\'e equations \cite{mmnp,mmreview}.

The perturbative series appearing in the $1/N$ expansion and in string theory have been comparatively less studied than their counterparts in 
Quantum Mechanics and Quantum Field Theory. One reason for this is the additional complexity of the problem, which involves an additional parameter: in the $1/N$ expansion, 
the coefficients are themselves functions of the 't Hooft coupling, while in the genus expansion of string theory, they are functions of $\alpha'$. In both cases the series 
are known to be asymptotic and to diverge factorially, like $(2g)!$ \cite{shenker}, but not much more is known about 
them. In some examples, like the bosonic string, it has been argued that the genus expansion is in general not Borel summable \cite{gp}. Limiting values of some 
scattering amplitudes of the bosonic string can be however resummed using the techniques of Borel resummation \cite{mo}. 
In some models of non-critical superstrings, the genus expansion is not Borel summable but there is a known 
non-perturbative completion \cite{kms}, and the structure one finds is similar to that of the double-well potential in Quantum Mechanics \cite{mmnp}. 
A recent attempt to resum the string perturbation series, by exploiting strong-weak coupling dualities, can be found in \cite{sen}.

Large $N$ dualities make it possible to relate the genus expansion of a string theory to the 't Hooft expansion of a gauge theory, 
and more importantly, they provide the non-perturbative objects behind 
these expansions. A particularly interesting example is the free energy of ABJM theory \cite{abjm} on a three-sphere, 
which depends on the rank $N$ of the gauge group and on the 
coupling constant $k$. It can be computed by localization \cite{kwy} and reduced to a matrix model which provides a concrete 
and relatively simple non-perturbative definition. 
The $1/N$ expansion of this matrix model is known in complete detail \cite{dmp} and can be generated in a recursive way. This gives us 
a unique opportunity to compare the asymptotic $1/N$ expansion of the gauge theory, as well as its Borel resummation, to the exact answer. 
By the AdS/CFT correspondence, the resulting $1/N$ series can be also regarded as the string perturbation series
 for the free energy of the type IIA superstring on AdS$_4 \times \IC\IP^3$ \cite{abjm}, and therefore we can address 
 longstanding questions on the nature of the string perturbation series by looking at this example.

In this paper we initiate a systematic investigation of these issues by using the techniques of Borel--Pad\'e resummation. 
As pointed out in \cite{dmpnp}, and in contrast to many previous examples, the perturbative genus expansion of the free energy of 
ABJM theory seems to be Borel summable. Hence, one can obtain accurate numerical values for the Borel--Pad\'e resummation of the series. 
However, we find strong evidence that the Borel resummation is {\it not} equal to the exact non-perturbative answer. This mismatch is controlled by 
{\it complex} instantons, which are known to exist in this theory and have been interpreted in terms of D2-brane instantons. This 
means that, in order to recover the exact answer, one should explicitly add to the Borel resummation of the perturbative series, the contributions due to these instantons. 

This result is somewhat surprising. On the one hand, there is no guarantee that the Borel resummation of a Borel summable series 
reconstructs the non-perturbative answer. There are sufficient conditions for this to be the case, like Watson's theorem and its refinements (see \cite{caliceti}), 
which typically require strong analyticity conditions on the underlying non-perturbative function. On the other hand, in most of the examples of Borel summable series in quantum 
theories, Borel resummation does reconstruct the correct answer, as in \cite{ggs}. As we will explain in this paper, the theory of resurgence 
suggests that this mismatch between the Borel resummation and the non-perturbative answer can be expected to happen in situations involving complex instantons. 
Indeed, an example of such a situation 
is the WKB series for the energies of the pure quartic oscillator, studied in \cite{bpv}. This series is asymptotic and oscillatory, 
and the leading singularity in the Borel plane is a complex instanton 
associated to complex trajectories in phase space. As already pointed out in \cite{bpv} (albeit in the context of a simpler resummation scheme known as 
optimal truncation), the contribution of this 
complex instanton has to be added explicitly, as in our string theory models. 
In this paper, in order to clarify the r\^ole of complex instantons, we revisit the quartic oscillator of \cite{bpv} in the context of Borel--Pad\'e resummation. 
We verify that, indeed, the difference between the Borel--Pad\'e resummation of the perturbative series and the exact answer 
is controlled by the complex instanton identified in \cite{bpv}. 

Our results lead to an important qualification concerning the non-perturbative structure of string theory. The standard lore is 
that string perturbation theory is not Borel summable, and therefore important non-perturbative effects have to be included \cite{gp}. Our results suggest that, 
even when string perturbation theory {\it is} Borel summable, additional non-perturbative 
corrections due to complex instantons might be required.  

 In this paper we also study a different, but closely related string perturbation series. 
The free energy of ABJM theory turns out to be related to the free energy of topological string theory on a toric Calabi--Yau 
manifold called local $\IP^1 \times \IP^1$ \cite{ dmp,mpabjm, mp}. In \cite{hmmo} this relationship was used to find a natural 
non-perturbative completion of the topological string free energy: the perturbative series of the topological string can be partially resummed 
by using the Gopakumar--Vafa representation \cite{gv2}, which has poles for an infinite number of values of the string coupling constant. In the non-perturbative completion 
of \cite{hmmo}, one adds to the Gopakumar--Vafa result non-perturbative corrections due to membrane instantons. It turns out that these corrections have poles 
which cancel precisely the divergences in the Gopakumar--Vafa resummation, and the final answer is finite (this is 
a consequence of the HMO cancellation mechanism of \cite{hmo2}). 

One could then ask at which extent the 
standard perturbative series of the topological string ``knows" about the non-perturbative completion proposed in \cite{hmmo}. It turns out that 
the Borel--Pad\'e resummation of the perturbative series does not reproduce the full non-perturbative answer, and the difference is 
due again to the presence of complex instantons in the theory. However, the Borel--Pad\'e resummation is smooth and does not display the singular behavior 
at the poles of the Gopakumar--Vafa representation. This indicates that the pole cancellation mechanism found in \cite{hmo2} and built in the proposal of \cite{hmmo} 
should be present in a non-perturbative completion of topological string theory. 

This paper is organized as follows. In section 2 we review the basic ideas and techniques of Borel resummation and the theory of resurgence used in this paper. We explain why 
complex instantons, although they do not obstruct Borel summability, might lead to relevant non-perturbative effects. We then show that this is exactly what happens in the example of the 
quantum-mechanical quartic oscillator studied in \cite{bpv, voros}. In section 3, we consider the resummation of the $1/N$ expansion in ABJM theory, and we compare in detail the results obtained 
in this way to the exact results. In section 4, we do a similar analysis for the genus expansion of a simple topological string model. Finally, in section 5 we list some conclusions and prospects for 
future explorations of this problem.

\sectiono{Borel resummation, non-perturbative effects, and the quartic oscillator}

\subsection{Borel resummation}

Our starting point is a formal power series of the form, 
\be
\label{aseries}
\varphi (z)=\sum_{n=0}^{\infty} a_n z^n. 
\ee
We will assume that the coefficients of this series diverge factorially, as
\be
a_n \sim A^{-n} n!.
\ee
In this case, the {\it Borel transform} of $\varphi$, which we will denote by $\widehat \varphi (\zeta)$, is defined as the series 
\be
\widehat \varphi (\zeta)=\sum_{n=0}^{\infty} {a_n \over n!} \zeta^n, 
\ee
and it has a finite radius of convergence $|A|$ at $\zeta=0$. We will sometimes refer to the complex plane of the variable $\zeta$ as the {\it Borel plane}. 
In some situations, we can analytically extend $\widehat \varphi(\zeta)$ to 
a function on the complex $\zeta$-plane. The resulting function will have singularities and branch cuts, but if it is analytic in a neighborhood 
of the positive real axis, and if it grows sufficiently slowly at infinity, we can define its Laplace transform
\be
\label{analytic}
s(\varphi) (z)=\int_0^{\infty}  \re^{-\zeta} \widehat \varphi (z \zeta) \, \rd \zeta =z^{-1} \int_0^{\infty} \re^{-\zeta/z} \widehat \varphi (\zeta) \, \rd \zeta, 
\ee
which will exist in some region of the complex $z$-plane. In this case, we say that the series $\varphi (z)$ is {\it Borel summable} and $s(\varphi) (z)$ is called the {\it Borel sum} or 
{\it Borel resummation} of 
$\varphi (z)$. 

In practice one only knows a few 
coefficients in the expansion of $\varphi (z)$, and this makes it very difficult to analytically continue the Borel transform to a neighbourhood of the 
positive real axis. A practical way to find accurate approximations to the resulting function is to use {\it Pad\'e approximants}. Given a series 
\be
g (z)=\sum_{k=0}^{\infty} a_k z^k
\ee
its Pad\'e approximant $[l/m]_g$, where $l, m$ are positive integers, is the rational function 
\be
\label{lmpade}
[l/m]_{g}(z) = {p_0 + p_1 z +\cdots + p_l z^l \over 
q_0 + q_1 z +\cdots + q_m z^m}, 
\ee
where $q_0$ is fixed to $1$, and one requires that
\be
g(z) -[l/m]_{g}(z) =\CO(z^{l+m+1}). 
\ee
This fixes the coefficients involved in (\ref{lmpade}).

The method of Pad\'e approximants can be combined with the theory of Borel transforms in the so-called Borel--Pad\'e method, which gives a very powerful tool to resum series. First, we use Pad\'e approximants to reconstruct the analytic continuation of the Borel transform $\widehat \varphi (z)$. There are various methods to do this, but one 
simple approach is to use the following Pad\'e approximant, 
\be
\label{paden}
\CP^{\varphi}_n(\zeta)= \Bigl[ [n/2]/[(n+1)/2] \Bigr]_{\widehat \varphi} (\zeta)
\ee
which requires knowledge of the first $n+1$ coefficients of the original series. The integral 
\be
\label{pade-approx}
 s(\varphi)_n(z)=z^{-1} \int_0^{\infty} \rd \zeta \, \re^{-\zeta/z}\CP^{\varphi}_n(\zeta)
 \ee
gives an approximation to the Borel resummation of the series (\ref{analytic}), which can be systematically improved by increasing $n$.

\begin{figure}[!ht]
\leavevmode
\begin{center}
\includegraphics[height=3.75cm]{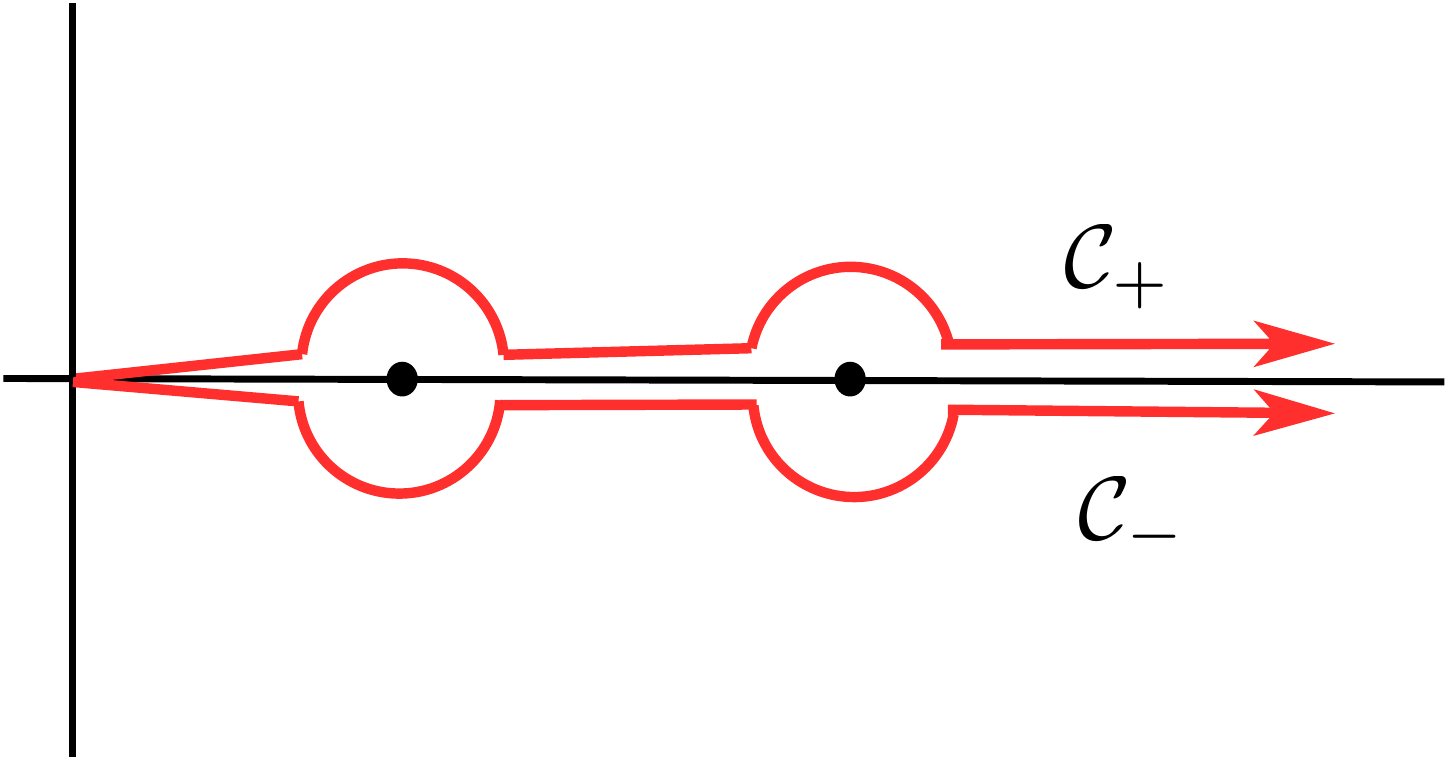}
\end{center}
\caption{The paths ${\cal C}_{\pm}$ avoiding the singularities of the Borel transform from above (respectively, below).}
\label{lateralfig}
\end{figure} 

Usually, the original asymptotic series is only the first term in what is called a {\it trans-series}, which takes into account all 
the non-perturbative sectors (see \cite{ss,mmreview} for 
reviews, \cite{igkm,mmnp,as} for developments of the general theory with applications to differential equations, 
\cite{du,cddu} for recent applications in Quantum Field Theory, and \cite{asv,cesv} 
for recent applications in string theory and matrix models). 
In its simplest incarnation, trans-series involve 
both the small parameter $z$ as well as the small exponentials 
\be
\re^{-S_\alpha/z}, \qquad \alpha \in \CA.
\ee
Here $\alpha \in \CA$ labels the different non-perturbative sectors of the theory. The trans-series is a formal infinite sum over all these sectors, of the form, 
\be
\label{sigz}
\Sigma(z)= \varphi (z) + \sum_{\alpha \in \CA} C_\alpha \re^{-S_\alpha/z} \varphi_{\alpha}(z), 
\ee
where the $\varphi_\alpha(z)$ are themselves formal power series in $z$, and $C_\alpha$ (in general complex numbers) are the weights of the instanton sectors. 
When the non-perturbative effects are associated to instantons, the quantities $S_\alpha$ are interpreted as instanton actions, and they usually appear as 
singularities in the complex plane of the Borel transform of $\varphi(z)$ (in simple cases, these actions are integer multiples
of a single instanton action $A$). In principle, the 
non-perturbative answer for the problem at hand can be obtained by performing a Borel resummation of the series $\varphi(z)$, $\varphi_\alpha(z)$, and then plugging in  
the result in (\ref{sigz}) with an appropriate choice of the $C_\alpha$. The resulting sum of trans-series is usually well-defined if $z$ is small enough.  
Since the Borel transforms of the formal power series appearing in (\ref{sigz}) might in general have singularities along the positive real axis, one has to consider 
as well lateral Borel resummations, 
\be
\label{lateralborel}
s_\pm (\varphi)(z)=z^{-1} \int_{\CC_{\pm}}  \re^{-\zeta/z} \widehat \varphi (\zeta)\, \rd \zeta, 
\ee
where the contours $\CC_{\pm}$ avoid the singularities and branch cuts by following paths slightly above or below the positive real axis, as in \figref{lateralfig}. 

\begin{figure}[!ht]
\leavevmode
\begin{center}
\includegraphics[height=3.75cm]{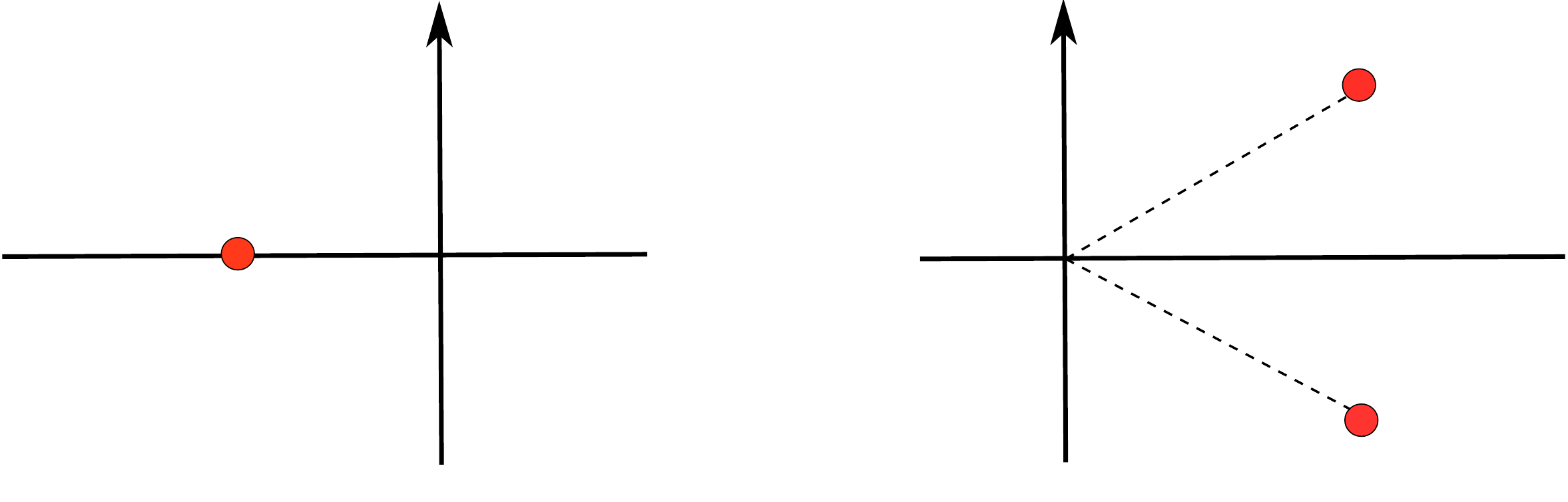}
\end{center}
\caption{Two different situations for Borel summability: in the case depicted on the left, the singularity occurs in the negative 
real axis of the Borel plane, but the corresponding instanton can not contribute to the final answer, since it would lead to an exponentially 
enhanced correction for $z>0$. However, in the case depicted on the right, we have two complex conjugate instantons whose actions 
have a real positive part. Although they do not obstruct Borel summability, they might lead to explicit 
non-perturbative corrections.}
\label{borel-sum-fig}
\end{figure} 
Let us suppose that we want to make sense of the formal trans-series for positive values of the argument $z>0$. It is clear that the terms in the 
trans-series where ${\rm Re}(S_\alpha)<0$ can not contribute to the final answer, since they are not exponentially suppressed when $z$ is small 
and positive, but rather exponentially enhanced. Thus, for 
example, if all the singularities in the Borel plane are real and negative, i.e. $S_\alpha<0$, we expect that the non-perturbative answer to 
the problem is simply given by the Borel resummation of the original perturbative series $\varphi(z)$. This is what happens for the energies of the 
quartic anharmonic oscillator: the only singularities of the Borel transform occur along the real negative axis \cite{ddp}, and the Borel 
resummation of the original perturbative series 
reconstructs the full answer \cite{ggs}. When some of the actions occur along the positive real axis, i.e. 
$S_\alpha>0$, they lead to obstructions to the Borel summability 
of the perturbative series $\varphi(z)$. One should then perform 
lateral Borel resummations and include explicitly the corresponding trans-series in 
(\ref{sigz}) in order to obtain the final answer. This is what happens in the case of 
the Hastings--McLeod solution to 
Painlev\'e II \cite{mmnp} and in the double-well potential in Quantum Mechanics \cite{zjj}. 

This discussion might lead to think that the only non-perturbative sectors that have to be included in (\ref{sigz}) are the ones associated to the obstruction of Borel summability, but this is not 
the case. Indeed, let us assume that, for some $\alpha \in \CA$, the corresponding action 
$S_{\alpha}$ is complex, but ${\rm Re}(S_\alpha)>0$. Clearly, the singularity at $S_\alpha$ does not obstruct Borel summability. 
However, since the corresponding term in (\ref{sigz}) is still exponentially small for small $z$, 
there is no {\it a priori} reason to exclude it from the final answer. We illustrate these considerations in \figref{borel-sum-fig}. 

Therefore, complex instantons whose actions have a real positive part 
play a subtle r\^ole in the theory of Borel resummation. It has been known for a long time that they lead to an oscillatory behavior 
in the large order asymptotics of the coefficients of the perturbative series \cite{dmpnp, blgzj,bpv,bdu}. What we would like to point out here is that, although 
they do not obstruct Borel summability, they might lead nonetheless to non-perturbative corrections, of order
\be
\CO\left( \re^{- {\rm Re}(S_\alpha)} \right).
\ee
 This was already observed in \cite{bpv}, albeit in a different language and in the context of optimal truncation. In addition, \cite{bpv} proposed a quantum-mechanical example 
 which displays this behavior in a non-trivial way: the quartic oscillator. We will now revisit this example in some detail, in order to exhibit the importance of complex instantons.

\subsection{The pure quartic oscillator}

The pure quartic oscillator is defined by the Hamiltonian (we follow the normalizations of \cite{bpv})
\be
H(q,p)= p^2 +V(q), \qquad V(q)=q^4. 
\ee
Since this is a confining potential, with $V(q) \rightarrow \infty$ as $q \rightarrow \infty$, the quantum Hamiltonian has a discrete spectrum of 
eigenvalues $E_k$. In this problem one can not use perturbation theory around the harmonic oscillator, and there are no parameters to play with (by elementary scaling, we have 
that $E_k(\hbar)= \hbar^{4/3} E_k(1)$). This suggests using the WKB expansion to find the energy levels, which is an asymptotic expansion for 
large quantum numbers. The starting point for this method is the well-known Bohr--Sommerfeld quantization condition, 
\be
{\rm vol}_0(E)=  2 \pi \hbar \left( k+\frac{1}{2} \right), \qquad k \ge 0. 
\ee
In this equation, 
\be
\label{volumeE}
{\rm vol}_0(E)= \oint_\gamma \lambda(q),
\ee
where $\gamma$ is a contour around the two real turning points defined by $V(q)=E$, and 
\be
\lambda(q)= p(q,E) \rd q, \qquad p(q,E) ={\sqrt{E -q^4}}
\ee
is a differential on the curve of constant energy defined by 
\be\label{spec}
H(q,p)=E. 
\ee
The notation (\ref{volumeE}) is due to the fact that the above integral computes the volume of phase space enclosed inside the curve (\ref{spec}). 

As it is well-known, the Bohr--Sommerfeld condition is just the leading term in a systematic $\hbar$ expansion. The 
quantum-corrected quantization condition, due to Dunham \cite{dunham}, can be formulated in a more geometric language as follows. We can solve the Schr\"odinger equation 
\be
\left(-\hbar^2 \frac{\dd^2}{\dd q^2}+q^4-E \right)\psi(q)=0
\ee
in terms of a function $p(q,E; \hbar)$ as
\be
\psi(q)= {1\over {\sqrt{ p(q,E; \hbar)}}} \exp \left( {\ri \over \hbar} \int^q p(q', E; \hbar) \rd q' \right). 
\ee
We then define the ``quantum" differential by 
\be
\lambda(q; \hbar)=p(q,E; \hbar) \rd q .
\ee
This has an expansion in powers of $\hbar$ whose first term is the ``classical" differential $\lambda(q)$. In terms of the ``quantum" differential $\lambda(q; \hbar)$ 
we define a quantum-corrected volume as 
\be
{\rm vol}_{\rm p}(E)=\oint_\gamma \lambda(q; \hbar), 
\ee
which reduces to ${\rm vol}_0(E)$ as $\hbar \rightarrow 0$. The Dunham quantization condition reads now, 
\be
\label{dunham-c}
{\rm vol}_{\rm p}(E)=2 \pi \hbar \left( k+\frac{1}{2} \right), \qquad k \ge 0. 
\ee

In the case of the pure quartic oscillator, one can write 
\be
{\rm vol}_{\rm p}(E)=\sum_{n=0}^{\infty}\hbar^{2n} \oint_{\gamma} u_{2n}(q) \rd q,
\ee
where $u_0(q)=p(q,E)$ and the higher order terms are given by the following recursion relation:
\be
\ba
u_{2n}&=(-1)^{n}v_{2n}, \qquad n\ge 0, \\
v_n&=\frac{1}{2p} \left( v_{n-1}'-\sum_{k=1}^{n-1}v_k v_{n-k} \right).
\ea
\ee
This implies that all the $u_{2n}(q)$ are sums of rational functions of the form $q^n/p^m$, and the contour integrals can be explicitly evaluated. By using the variable
\be
\label{sigE}
\sigma = \frac{\Gamma(1/4)^2}{3 \hbar}\sqrt{\frac{2}{\pi}}E^{3/4},
\ee
one finds that
\be \label{energyseries1}
{1\over \hbar} {\rm vol}_{\rm p}(E)=  \sum_{n \geq 0} b_n \sigma^{1-2n},
\ee
where the coefficients $b_n$ can be computed in closed form, and $b_0=1$. 

The series appearing in (\ref{energyseries1}) has 
zero radius of convergence, therefore we should expect a rich non-perturbative structure in the theory. 
Such a structure has been studied in detail in \cite{bpv,voros}. The first step in understanding non-perturbative aspects of this model 
is to look for all possible saddle-points of the path integral. Since the quartic potential has a single minimum, one could think that the only 
saddle-point is the classical trajectory of energy $E$ between the turning points $\pm E^{1/4}$, which corresponds to the cycle $\gamma$ in (\ref{volumeE}). However, 
one should look for general, {\it complex} saddle-points\footnote{Strictly speaking, instantons are just a particular class of such saddle-points, describing a trajectory in real space but in imaginary 
or Euclidean time. However, we will refer to a general saddle-point of the complexified theory also as an instanton configuration.}. 
The curve (\ref{spec}), once it is complexified, describes a Riemann surface of genus one with a lattice of one-cycles. The perturbative quantization condition (\ref{dunham-c}) 
involves the cycle going around the two real turning points, but there are other one-cycles related to non-trivial, complex saddle-points. The existence of these 
cycles can be seen very explicitly by looking at complexified classical trajectories. The classical solution to the EOM with energy $E$ is given by the trajectory 
\be
\label{class-tr}
q(t)= E^{1/4} {\rm cn}\left( 2 {\sqrt{2}} E^{1/4} t, {1\over {\sqrt{2}}}  \right), 
\ee
where we have fixed one integration constant due to time translation invariance. 
Here, ${\rm cn}(u, k)$ is a Jacobi elliptic function. As noticed in \cite{bpv}, this function has a complex lattice $\Lambda$ of periods in the $t$ plane, generated by 
\be
T_1= {T + \ri T \over 2}, \qquad T_2= -{T-\ri T \over 2}, 
\ee
where
\be
T={\sqrt{2}} E^{-1/4} K\left(1/{\sqrt{2}} \right)
\ee
is the real period of the classical trajectory (\ref{class-tr}), and $K(k)$ is the elliptic integral of the first kind. 
Any period in $\Lambda$ leads to a complex, periodic trajectory. The trajectories with periods $T_1$, $T_2$ 
go around the real turning point $-E^{1/4}$ and the complex turning points $\pm \ri E^{1/4}$, respectively. They have actions $\sigma S_{1,2}$, where
\be
S_1= {1+ \ri \over 2}, \qquad S_2= -{1 -\ri \over 2}. 
\ee
Therefore, the actions of the complex trajectories associated to the periods in $\Lambda$ are given by $\sigma$, times
\be
\label{toutes}
n S_1+ m S_2, \qquad n, m \in \IZ. 
\ee
The cycle $\gamma$ appearing in (\ref{volumeE}) corresponds to the real trajectory (\ref{class-tr}) with period $T$, and it is associated to the point $S_1-S_2=1$, with action $\sigma$. 
The lattice of points (\ref{toutes}) gives the possible singularities of the Borel transform of the series (\ref{energyseries1}), 
which we define as follows: we write the perturbative series (\ref{energyseries1}) 
as
\be 
{1\over \hbar} {\rm vol}_p (E)=\sigma +
{1\over \sigma} \varphi(\sigma), \qquad \varphi(\sigma)= \sum\limits _{n\geq0}b_{n+1} \sigma^{-2n}.
\ee
The Borel transform is then defined by
\be 
\label{borelb}
\widehat \varphi(\zeta)= \sum_{n\geq0} {b_{n+1} \over (2n)!} \zeta^{2n}. 
\ee
This definition is slightly different from the one used in \cite{voros}, but leads to the same structure of singularities in the Borel plane. 
However, not all the instanton actions in (\ref{toutes}) lead to {\it actual} singularities in the Borel transform. These have been 
determined in \cite{voros}, and they are given by the points $n S_1$ and $n S_2$, where $n \in \IZ\backslash\{ 0\}$, as well as by the points
\be
n (S_1- S_2), \qquad n \in \IZ\backslash\{ 0\}. 
\ee
The singularities which are closest to the origin are $\pm S_1$ and $\pm S_2$, and they correspond to complex saddles with actions 
\be
{\pm 1 \pm \ri \over 2} \sigma. 
\ee
Through the standard connection to the large order behavior of the perturbative series, they lead to an oscillatory behavior 
for the coefficients $b_n$ \cite{bpv}. Note as well that there is an infinite number of singularities along the positive real axis, and the closest one to the origin 
occurs at $S_1-S_2=1$. This clearly leads to an obstruction to Borel summability. However, this obstruction comes from a sub-dominant singularity, and it is only seen in exponentially small, 
subleading corrections to the large order behavior of the coefficients $b_n$ \cite{voros}.  

\begin{figure}
\begin{center}
\includegraphics[scale=0.75]{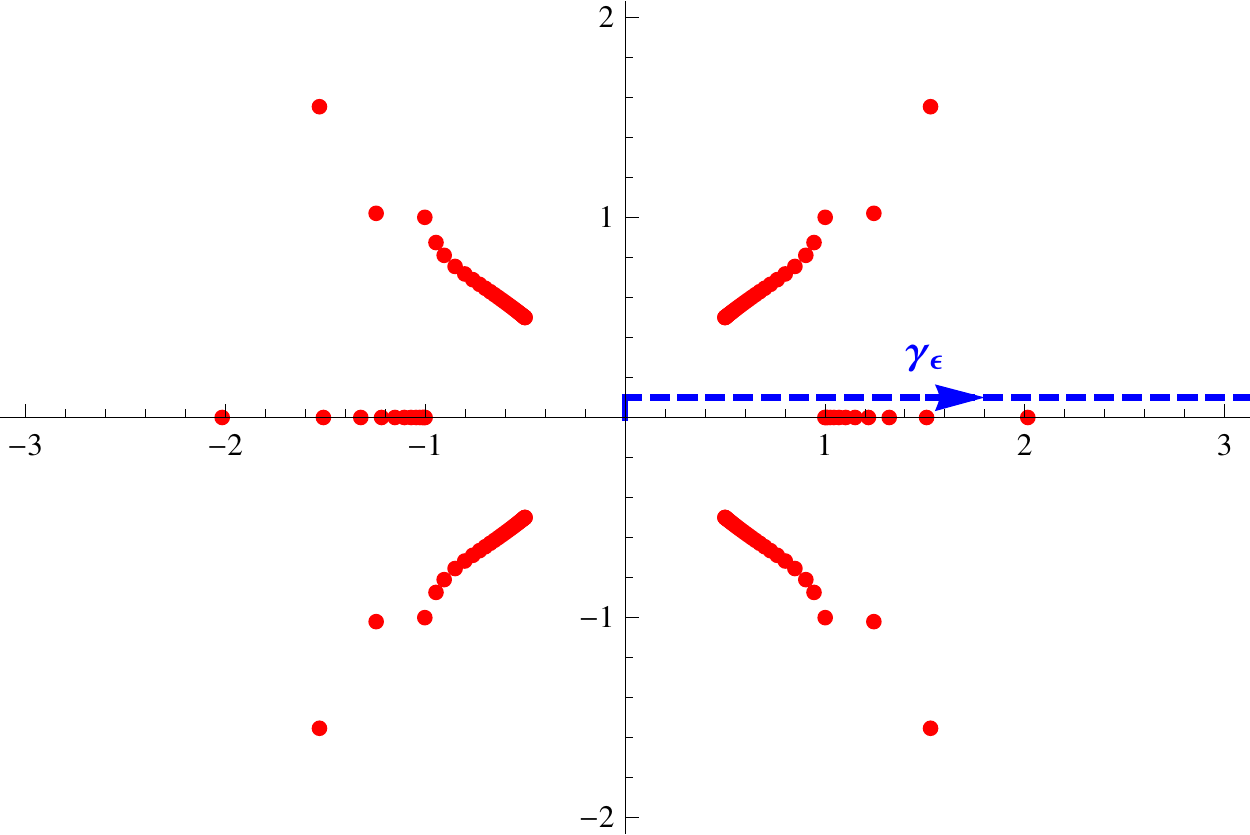}
\caption{The poles of the Pad\'e approximant (\ref{paden}) in the $\zeta$ plane, with $n=320$, for the series (\ref{borelb}).}
 \label{polemapbn}
\end{center}
\end{figure}

We would like to compare the results for the energy spectrum obtained by Borel--Pad\'e resummation of the series (\ref{energyseries1}), to the values obtained by solving the Schr\"odinger 
equation, as in \cite{exactAnharmonic2, exactAnharmonic1}. Since there is a singularity in the positive real axis of the Borel plane, 
we have to consider lateral Borel--Pad\'e resummations. In order to do this, 
we have first generated a large number of terms in the series (\ref{energyseries1}), and then computed Pad\'e approximants (\ref{paden}) of the 
Borel transform (\ref{borelb}), for different values of $n$. 

The first piece of information that can be extracted from the Pad\'e approximants is the structure of singularities in the Borel plane. 
The Pad\'e approximants are by construction rational functions, therefore their only singularities are poles. 
However, the accumulation of their poles along segments signals the presence of branch cut singularities in the Borel transform. 
The pole structure of the Pad\'e approximant, for $n=320$, is shown in \figref{polemapbn}. 
The accumulation of the poles gives a good approximation to the analytic structure of the Borel transform found by Voros 
(compare our \figref{polemapbn} with Fig. 24 in \cite{voros}). The four lines at angles $k \pi/4$, with $k=\pm 1, \pm 3$, 
start at the points $\pm S_1$, $\pm S_2$ and signal the presence of the 
complex instantons. 

As we can see in \figref{polemapbn}, there is also an accumulation of poles of the Pad\'e approximant in the positive real axis, starting at $\zeta=1$, 
which signals the subleading, real instanton at $S_1-S_2$. The lateral Borel resummation is performed along the path 
\be
\label{cip}
\gamma_\epsilon=L [0, \ri \epsilon]+ L[\ri \epsilon, \infty +\ri \epsilon)
\ee
where $L[a,b]$ denotes a straight line in the complex plane from $a$ to $b$ (in practice, we take $\epsilon \approx 10^{-2}$). This path, 
which is also shown in \figref{polemapbn}, avoids the poles in the positive real axis, and it is homotopic 
to the path $\CC_+$ shown in \figref{lateralfig}. 
After lateral resummation, we produce a function 
\be
\label{fnp-s}
F_n^{\rm BP}(\sigma)=\sigma + {1\over \sigma} s(\varphi)_{n, \gamma_\epsilon} (\sigma),
\ee
where
\be
 s(\varphi)_{n, \gamma_\epsilon} (z)=z^{-1} \int_{\gamma_\epsilon} \rd \zeta \, \re^{-\zeta/z}\CP^{\varphi}_n(\zeta)
 \ee
 and $n$ is the order of the Pad\'e approximant. The functions (\ref{fnp-s}), for $n=1, 2, \cdots$, provide approximations to the Borel resummation of the series (\ref{energyseries1}). They are 
 however complex, due to the complex integration path (\ref{cip}). One can verify that 
\be
{\rm Im}\left( F^{\rm BP}_n(\sigma) \right)\sim \exp\left( -\sigma\right), 
\ee
for sufficiently large $n$ (we verified it explicitly for $n=95$). This reflects the ambiguity associated to the real instanton obstructing Borel summability. 

We can now find a numerical approximation to the energy levels by 
using the quantization condition based on the Borel resummation of (\ref{energyseries1}), i.e. we use 
\be
\label{num-en}
{\rm Re} \left( F^{\rm BP}_n(\sigma) \right)= 2 \pi  \left(k +{1\over 2} \right), \qquad k \ge 0. 
\ee
We will denote by $E_n ^{(0)}(k)$ the energy levels obtained from this quantization condition 
(the superscript $(0)$ indicates that we are not considering instanton contributions explicitly). Some of the results of our calculation 
are displayed in Table \ref{Etable}. 

\begin{table}[h!] \begin{center}
 \begin{tabular}{|c|c|c|c|} \hline
	
	 $k$ & $E(k)$ & $E_{320}^{(0)} (k)$  \\ \hline
	  6 &26.528 471 183 682 518 191 8&26.528 471 181 399 704 803   \\ \hline
	  3 & 11.644 745 511 378 & 11.644 768 005 3  \\ \hline
	  0 &  1.060  & 0.96  \\ \hline
	
     \end{tabular}
        \caption{Energies of the levels $k=0,3,6$ of the quartic oscillator with $\hbar=1$. The first column shows the value obtained numerically in \cite{exactAnharmonic2} for $k=0,6$ and \cite{exactAnharmonic1} for $k=3$, while the 
        second column shows the value of the energy obtained by using the quantization condition (\ref{num-en}) with $n=320$. All the given digits are stable.} \label{Etable}
        \end{center}
   \end{table}

Since we are not including the effects of the real instanton, we know that the energy levels obtained with the above procedure 
will have an accuracy not better than $\re^{-\sigma}$, where $\sigma$ is given by (\ref{sigE})\footnote{Since we are imposing the quantization condition (\ref{dunham-c}), 
$\sigma$ is, for large $k$, of order $2 \pi k$.}. However, by comparing the real 
part of the Borel--Pad\'e resummation to 
the energy levels calculated numerically, one notices that the disagreement is not of order 
$\re^{-\sigma}$, but rather of order 
$\re^{-\sigma/2}$. This was already pointed out in \cite{bpv}, in the context of optimal truncation. In that paper, Balian, Parisi and Voros argued that one should correct the perturbative 
WKB quantization condition (\ref{dunham-c}) by adding explicit, non-perturbative contributions to the volume of phase space, which incorporate the effect of complex instantons. These corrections 
are due to the complex instantons associated to the points $S_1$, $-S_2$, whose action has a real part given by $\sigma/2$. The leading term of their
contribution was determined in \cite{bpv}, and it is given by 
\be
\label{one-instanton}
{\rm vol}_{\rm np}(E)= \mp 2 \,  {\rm arctan} \left[ \exp \left( \frac{\ri} {2\hbar}   \oint_{\gamma'} \lambda(q; \hbar) \right) \right] +\cdots,
\ee
where the $\mp$ sign corresponds to wavefunctions with even (respectively, odd) parity, and 
$\gamma'$ is  a contour in the complex $q$-plane around the two imaginary turning points $\ri E^{1/4}$, $-\ri E^{1/4}$. 
Note that (\ref{one-instanton}) is just the leading non-perturbative correction to the volume of phase space. There should be additional non-perturbative corrections coming from higher instantons, which are exponentially suppressed as compared to (\ref{one-instanton}). 

We can now incorporate the above non-perturbative correction to the quantization condition
\be
{\rm vol}(E)= {\rm vol}_{\rm p}(E)+ {\rm vol}_{\rm np}(E)= 2 \pi \hbar \left( k+{1\over 2} \right). 
\ee
This correction involves an additional asymptotic series, coming from the integration along the cycle $\gamma'$. It turns out that this series can be expressed in terms of the 
same coefficients $b_n$ \cite{bpv}, namely, 
\be
 \label{energyseries2}
\frac{\ri} {2\hbar}   \oint_{\gamma'} \lambda(q; \hbar) = -\frac{1}{2} \sum_{n \geq 0} (-1)^{n}b_n \sigma^{1-2n},
\ee
and it can be also analyzed with the Borel--Pad\'e resummation method. In this case there are no singularities in the positive real axis \cite{voros}, 
and we can do a standard Borel--Pad\'e resummation. If we proceed exactly as we did for (\ref{energyseries1}), we obtain, from the formal series (\ref{energyseries2}), 
a function $G_n ^{\rm BP}(\sigma)$, where $n$ is again the order of the Pad\'e approximant as defined in (\ref{paden}). 

We can now improve the calculation of the energies by studying the quantization condition 
\be
\label{one-inst}
{\rm Re} \left( F^{\rm BP}_n(\sigma) \right) -2 (-1)^k \, {\rm arctan}\left(G_n ^{\rm BP}(\sigma) \right) = 2 \pi  \left(k +{1\over 2} \right), \qquad k \ge 0. 
\ee
We will denote by $E_n ^{(1)}(k)$ the energy levels obtained from this quantization condition. Once we add this correction, the resulting energies are much 
closer to the numerical values, see for example the results in 
Table \ref{Etable2}. Of course, the values $E_n ^{(1)}(k)$ still differ from the exact ones by higher instanton corrections. It should be noted that our calculation of the energies, 
by using the Borel--Pad\'e method, improves the results of \cite{bpv}, which were obtained by doing optimal truncation 
in both asymptotic series, (\ref{energyseries1}) and (\ref{energyseries2}). 
\begin{table}[h] \begin{center}
\begin{small}
 \begin{tabular}{|c|c|c|} \hline
	
	 $k$ & $E(k)$ & $E_{320}^{(1)} (k)$    \\ \hline
	  6 &26.528 471 183 682 518 191 813 828 183  &26.528 471 183 682 518 191 813 828 183   \\ \hline
	  3 & 11.644 745 511 378 & 11.644 745 511 378  \\ \hline
	  0 &  1.060 4 & 1.060 4  \\ \hline
	
     \end{tabular}
\end{small}
        \caption{Energies of the levels $k=0,3,6$ of the quartic oscillator with $\hbar=1$. The first column shows the value obtained numerically in \cite{exactAnharmonic2} for $k=0,6$ and \cite{exactAnharmonic1} for $k=3$, while the 
        second column shows the value of the energy obtained by using the quantization condition (\ref{one-inst}) with $n=320$. Only the stable digits are given.} \label{Etable2}
        \end{center}
   \end{table}

It has been proposed in \cite{jentschura} that, when there are complex poles in the Pad\'e approximant, one should add to the Borel--Pad\'e resummation the residues of these poles. 
This indeed adds exponentially small corrections with the required magnitude (in this case, of order $\exp(-\sigma/2)$). We have verified that including these residues leads to an increased 
accuracy in the numerical values of the energy levels, similar to the one obtained by considering the one-instanton correction (\ref{one-instanton}). It would be interesting to analyze 
this in more detail, but in any case the prescription of \cite{jentschura} does not seem to be universally valid, and it does not lead to an improvement of 
the approximation for the string perturbation series which we will analyze in subsequent sections. 

The main conclusion of our analysis of the pure quartic oscillator is that Borel summability is {\it not} enough to reconstruct a non-perturbative answer from 
the resummed perturbative series: complex instantons, which are not an obstruction to the Borel resummation, have to be nevertheless included explicitly in the full answer. 
Of course, the quartic oscillator is, technically speaking, not Borel summable, but this is a not crucial issue, since the instantons which are responsible for the breakdown of Borel 
summability lead to sub-dominant non-perturbative effects. The dominant non-perturbative effects are due to complex instantons. As we will see, 
in the case of the string perturbation series of ABJM theory, the situation is even more transparent: the series seems to be Borel 
summable (no poles accumulate in the positive real axis), yet there are non-perturbative effects associated to complex instantons which should be 
included explicitly.

\sectiono{Resumming the $1/N$ expansion in ABJM theory}

In this section we will combine the AdS/CFT correspondence for ABJM theory \cite{malda,abjm} with the results on the $1/N$ expansion of the matrix model 
computing its free energy on the the three-sphere \cite{kwy}, to obtain quantitative results on the resummation of the string perturbation series. We should note that the 
matrix model of \cite{kwy} was used in \cite{russo} to analyze the perturbative series in $1/k$ for the free energy, at fixed $N$. This is a one-parameter problem. Here, in contrast, 
we study the $1/N$ expansion, in which each coefficient is itself a non-trivial function of the 't Hooft parameter. 

ABJM theory \cite{abjm,abjmreview} is a conformally invariant, Chern--Simons--matter theory in 
three dimensions with gauge group $U(N)_k \times U(N)_{-k}$ and $\CN=6$ supersymmetry. 
The Chern--Simons actions for the gauge groups have couplings $k$ and $-k$, respectively. The theory contains as well four hypermultiplets 
in the bifundamental representation of the gauge group. The 't Hooft parameter of this theory is defined as
\be\lambda={N\over k}.
\ee
In \cite{kwy} it was shown, through a beautiful application of localization 
techniques, that the partition function of ABJM theory on the three-sphere can be computed by a 
matrix model (see \cite{mmcslectures} 
for a pedagogical review). This matrix model is given by 
\be
\label{kapmm}
Z(N, k)={1\over N!^2} \int \prod_{i=1}^{N}{ \rd \mu_i \rd \nu_j  \over (2 \pi)^2} {\prod_{i<j} \sinh^2 \left( {\mu_i -\mu_j \over 2}\right)  \sinh^2 \left( {\nu_i -\nu_j \over 2}\right) \over 
\prod_{i,j}  \cosh^2 \left( {\mu_i -\nu_j \over 2}\right)} \re^{{ \ri k \over 4 \pi}\left(  \sum_i \mu_i^2 -\sum_j \nu_j^2\right)}.
\ee
The free energy, defined as $F(N, k)= \log Z(N,k)$, has a $1/N$ expansion 
of the form 
\be
\label{fgs}
F(\lambda, k)=\sum_{g=0}^{\infty} \left( {2 \pi \over k}\right)^{2g-2} F_g(\lambda).
\ee
ABJM theory has been conjectured to be dual to type IIA superstring theory on AdS$_4\times \IC\IP^3$. This theory has two parameters, the string 
coupling constant $g_{\rm st}$ and the radius $L$ of the AdS space, and they are related to the parameters $\lambda$, $k$ of ABJM theory by 
\be
\label{ssdic}
\ba
k^2&=g_{\rm st}^{-2} \left( {L \over \ell_s}\right)^2,\\
\lambda-{1\over 24}&={1\over 32 \pi^2 } \left( {L \over \ell_s}\right)^4 \left( 1-{4 \pi^2 g_{\rm st}^2 \over 3} \left( {\ell_s \over L}\right)^6 \right), 
\ea
\ee
where $\ell_s$ is the string length. Here, we have used the corrected dictionary proposed in \cite{bh,ahho}, although our results will not depend on its details. According to the AdS/CFT correspondence, the 
free energy (\ref{fgs}) is the free energy of type IIA superstring theory on the AdS background, and its $1/N$ expansion (\ref{fgs}) corresponds to the genus expansion of 
the superstring. 

The genus $g$ free energies appearing in (\ref{fgs}) 
were determined in \cite{dmp} by using various techniques. The strong coupling regime of the free energies at genus zero and one reproduces the expected answer from supergravity \cite{dmp,bgms}. 
We will now review the structure of these free energies. Their natural variable is the parameter $\kappa$, which is related to the 't Hooft coupling by \cite{mpabjm,dmp}
 \be
 \label{lamkap}
 \lambda(\kappa)={\kappa \over 8 \pi}   {~}_3F_2\left(\frac{1}{2},\frac{1}{2},\frac{1}{2};1,\frac{3}{2};-\frac{\kappa^2
   }{16}\right). 
\ee
The genus zero free energy is determined by the equation, 
\be
\label{comf}
- \partial_\lambda F_0={\kappa \over 4} G^{2,3}_{3,3} \left( \begin{array}{ccc} {1\over 2}, & {1\over 2},& {1\over 2} \\ 0, & 0,&-{1\over 2} \end{array} \biggl| -{\kappa^2\over 16}\right)+{ \pi^2 \ri \kappa \over 2} 
  {~}_3F_2\left(\frac{1}{2},\frac{1}{2},\frac{1}{2};1,\frac{3}{2};-\frac{\kappa^2 
   }{16}\right),
\ee
where $G^{2,3}_{3,3}$ is a Meijer function\footnote{The free energies used in this paper have an overall factor $(-1)^{g-1}$ w.r.t. the 
ones used in \cite{dmp,dmpnp}.}. 
The integration constant can be fixed by looking at the weak coupling limit \cite{dmp, hanada}. For $g\ge 1$, the free energies are quasi-modular forms with 
modular parameter
\be
\label{tauex}
\tau=\ri  {K'\left({\ri \kappa \over 4}\right)\over K \left({\ri \kappa \over 4}\right)}.
\ee
For $g=1$, one has 
\be
F_1=-\log \, \eta (\tau)+2\zeta'(-1)+{1\over 6}\log \left({ \ri \pi \over 2 k}\right),
\ee
where $\eta$ is the usual Dedekind eta function. For $g\ge 2$, the $F_g$s can be written in terms of $E_2(\tau)$ (the standard Eisenstein series), 
$b(\tau)$ and $d(\tau)$, where 
\be
b(\tau)=\vartheta_2^4(\tau), \qquad d(\tau)=\vartheta_4^4(\tau),
\end{equation} 
are standard Jacobi theta functions. More precisely, they have the general structure  
\be
F_g(\lambda)={1\over \left( b(\tau) d^2(\tau) \right)^{g-1}} \sum_{k=0}^{3g-3}
E_2^{k}(\tau) p^{(g)}_k\left(b(\tau),d(\tau)\right), \qquad g\ge 2, 
\ee
where $p^{(g)}_k\left(b(\tau),d(\tau)\right) $ are polynomials in $b(\tau)$, $d(\tau)$ of modular weight $6g-6-2k$. The genus $g$ free energies $F_g(\lambda)$ obtained 
in this way are exact functions of the 't Hooft parameter, and they provide interpolating functions between the weak and the strong coupling regimes.  
 
The nature of the series (\ref{fgs}) was investigated in \cite{dmpnp}. As usual in string theory and in the $1/N$ expansion, at fixed $\lambda$, the genus $g$ free energies diverge factorially \cite{shenker}, 
\be
F_g(\lambda) \sim (A(\lambda))^{-2g} (2g)!. 
\ee
A first question one can ask is: what are the possible instanton actions appearing in the non-perturbative sector, and how do they manifest themselves in the 
large order behavior of the free energies? In \cite{dmpnp}, based on previous work on instantons in matrix models (reviewed in for example \cite{mmreview}), a proposal 
was made for the instanton actions. The large $N$ limit of the matrix model (\ref{kapmm}) is controlled by a spectral curve of genus one, and there is a 
lattice of periods, just as in the case of the quartic oscillator analyzed in the previous section. In addition, there is a constant period. 
The conjecture of \cite{dmpnp} is that the instanton actions are linear integer combinations of the two independent periods of the spectral curve, and of the 
constant period. This proposal is very much along the lines of \cite{bpv}, since they both involve the periods of a complexified curve.

The proposal of \cite{dmpnp} can be checked by looking at the large order behavior of the series (\ref{fgs}). 
At large $\lambda$, the leading behavior of $F_g$ is dominated by the so-called constant map contribution, 
\be
\label{cmap}
F_g(\lambda) = c_g + \CO\left( \lambda^{3/2-2g}\right), \qquad g\ge 2,
\ee
where
\be
c_g=  {4^{g-1} (-1)^g |B_{2g} B_{2g-2}| \over g (2g-2) (2g-2)!}, 
\ee
and $B_{2g}$ are Bernoulli numbers. The large order behavior of these coefficients is controlled by the constant period \cite{ps,mmreview}
\be
\label{cm-inst}
A=2 \pi^2 \ri,
\ee
together with its complex conjugate. They lead to a pair of complex conjugate singularities in the Borel plane. However, 
this gives the ``trivial" part of the asymptotic behavior. 
Much more interesting is the subleading singularity, which can be obtained from the study of the large 
order behavior of the sequence $F_g(\lambda) - c_g$ at sufficiently large $\lambda$ (in practice, $\lambda \gtrsim 0.75$). It is controlled by the action 
\be
\label{s-inst}
A_s(\lambda) = -{1\over \pi}  \partial_\lambda F_0 + \ri \pi^2, 
\ee
which is one of the periods of the spectral curve. Since this action is complex, it can be written as 
\be
A_s (\lambda)= \left|A_s(\lambda)\right| \re^{\ri \theta_s(\lambda)}, 
\ee
and it leads to an oscillatory behavior in the sequence $F_g(\lambda)-c_g$:
\be
\label{fgcg}
F_g(\lambda) - c_g \sim  \left| A_s(\lambda)\right| ^{-2g} \cos\left( 2g\theta_s (\lambda) +\delta_s(\lambda) \right)(2g)!, 
\ee
where $\delta_s(\lambda)$ is an unknown function of $\lambda$. The behavior (\ref{fgcg}) was tested 
numerically in \cite{dmpnp}. As explained in \cite{dmpnp}, for smaller values of $\lambda$, the dominant action is no longer (\ref{s-inst}), but 
\be 
\label{w-inst} 
A_w(\lambda)= 4 \ri \pi ^2 \lambda.
\ee
In addition, a study of the lattice of the periods in \cite{dmpnp} led to the conclusion that 
there are no singularities on the positive real axis of the Borel plane. As we will see in a moment, our numerical results for the Borel--Pad\'e transform 
seem to confirm the Borel summability of the asymptotic series (\ref{fgs}). 

The results of \cite{dmpnp} open the window to an analysis of the Borel--Pad\'e resummation of (\ref{fgs}). Using the techniques of \cite{dmp}, we can 
generate the free energies in (\ref{fgs}) up to genus $30$. As we will see, this gives already good numerical results. 
On the other hand, since (\ref{fgs}) is the $1/N$ expansion of the matrix model 
(\ref{kapmm}), we know what is the non-perturbative object that we should compare to this resummation: the free energy of 
the matrix model $F(N,k)$ for finite values of $N$ and $k$\footnote{In \cite{mmnp}, 
a similar study was performed, in which the $1/N$ expansion of the Gross--Witten--Wadia model 
was resummed and compared to the exact non-perturbative answer. In this model, the genus expansion is not Borel summable and the 
instanton sectors which should be included are well understood.}. It turns out that the 
partition function $Z(N,k)$ has been computed analytically with the TBA equations of \cite{zamo,tw} in 
\cite{py,hmo1,hmo2}, for various integer values of $N$ and $k$. In particular, \cite{hmo2} 
gives results for $k=1,2,3,4,6$ and $N=1,2,...,N_{{\rm max},k}$ where $N_{{\rm max},(1,2,3,4,6)}=(44,20,18,16,14)$.

To proceed with the Borel--Pad\'e resummation, we write (\ref{fgs}) as 
\be 
F(\lambda, z)=z^{-2} F_0(\lambda)+F_1(\lambda)+\sum_{g\geq 2} z^{2g-2}F_g(\lambda), 
 \ee
 where
 \be
 z={2 \pi \over k}. 
 \ee
Then we consider the Borel transform of
\be
\varphi (z)= \sum_{g\geq 2} F_g (\lambda)z^{2g-2}, 
\ee
which is given by 
\be
\label{btfg}
\widehat \varphi (\zeta)=\sum_{g\geq 2} {F_g(\lambda)\over (2g-2)!}\zeta ^{2g-2}, 
\ee 
and we fix the value of $\lambda$ to obtain a numerical series. We consider Pad\'e approximants of order $n$ for the Borel transform, as in (\ref{paden}). 

\begin{figure}
\begin{center}
\includegraphics[scale=0.5]{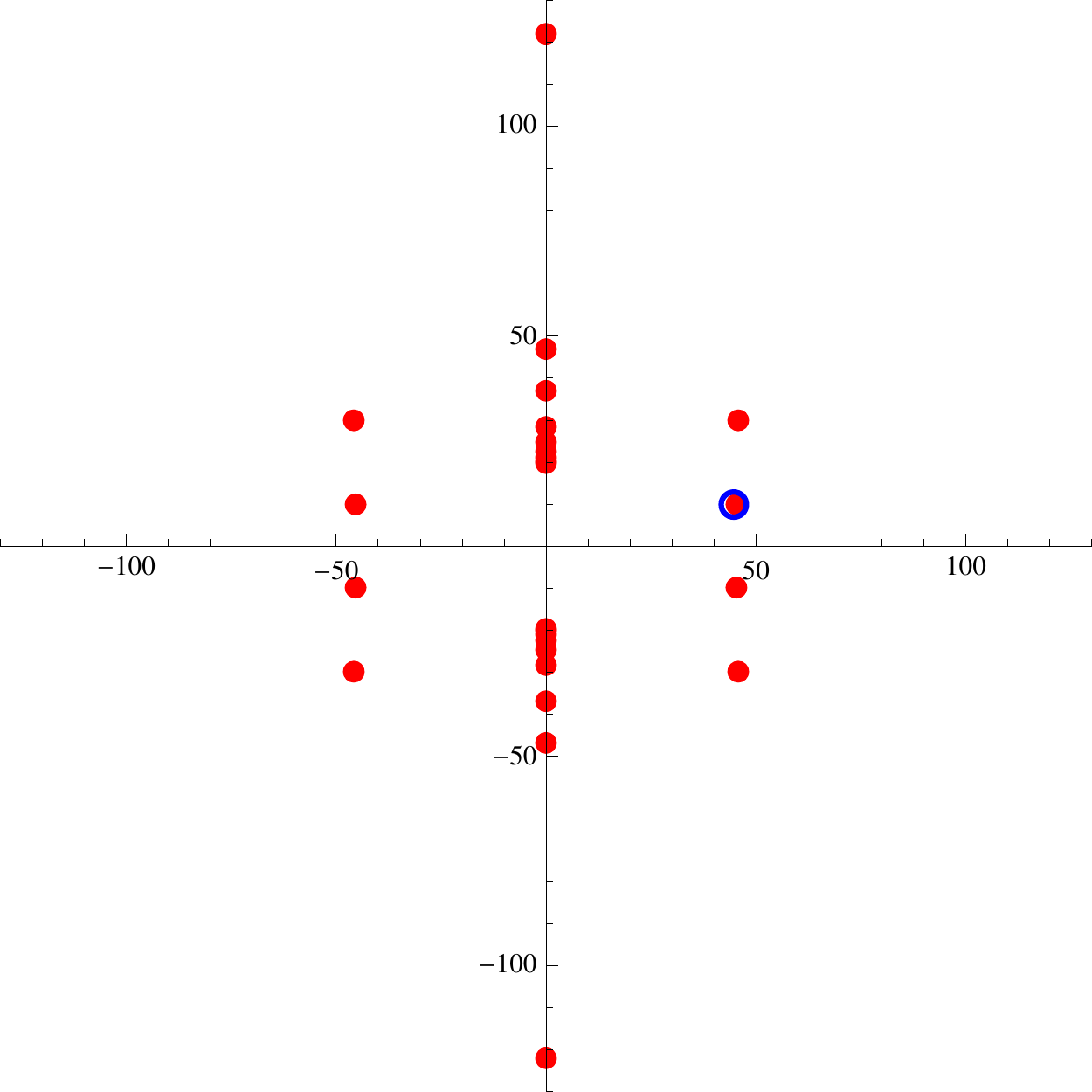} \qquad \qquad \includegraphics[scale=0.5]{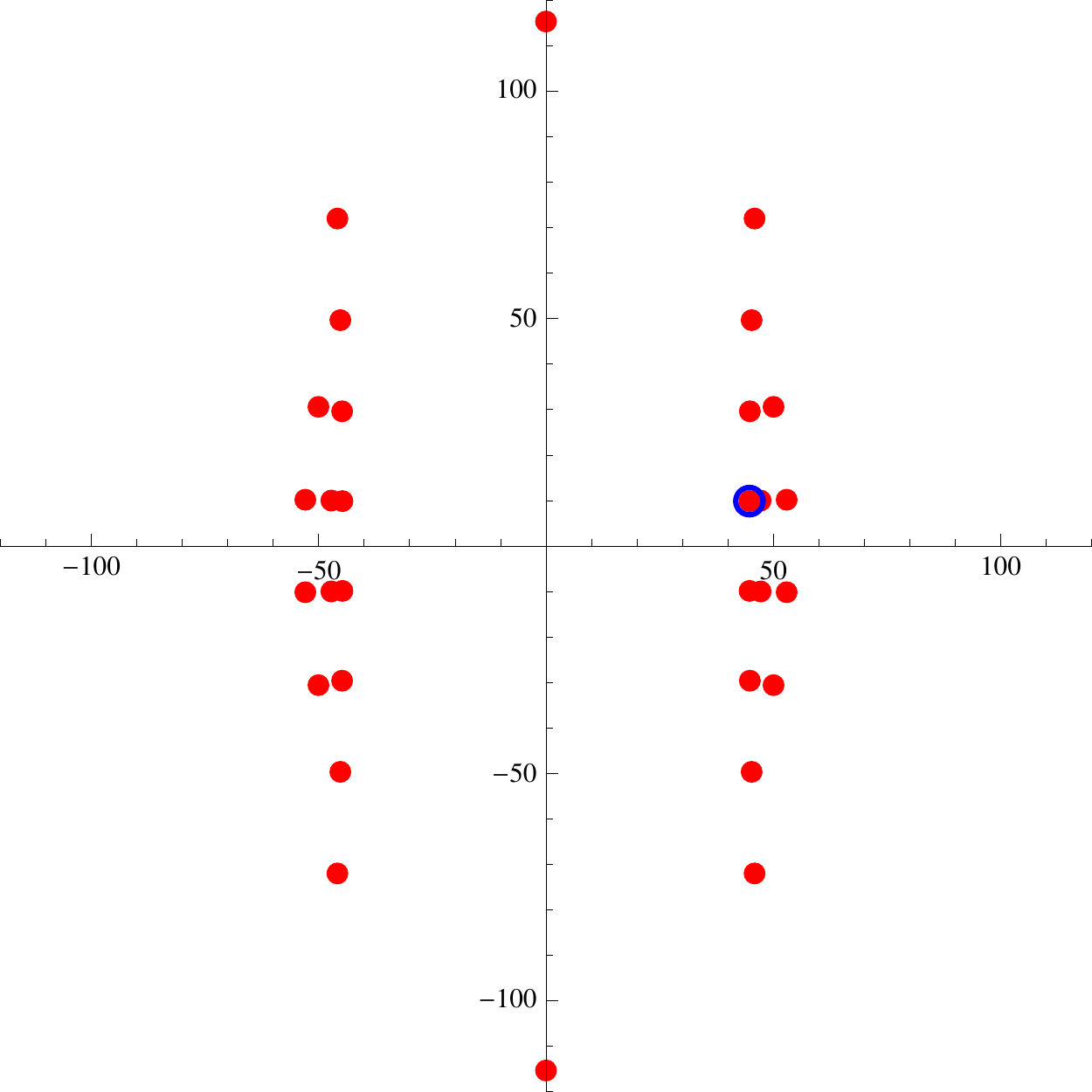} 
\caption{The red dots signal the location of the 
poles of the Pad\'e approximant (\ref{paden}), for the series (\ref{btfg}) with $\lambda\approx 2.61$. 
In the figure on the left we have included the full $F_g(\lambda)$, while in the figure on the right 
we have subtracted the contribution of constant maps. The blue circle corresponds to the numerical value of the complex instanton action 
$A_s(\lambda)$. The degree of the Pad\'e approximant is $n=54$ (left) 
and $n=60$ (right).}
\label{polesPadeOr}
\end{center}
\end{figure}

As we explained in the example of the quartic oscillator, the first information we can obtain from the Borel--Pad\'e transform is the singularity structure in the Borel plane of 
the $\zeta$ variable. This can be studied by looking at the poles of the Pad\'e approximants. 
We show the location of these poles in \figref{polesPadeOr} for $\lambda \approx 2.61$, where we consider 
both the Pad\'e approximant of the series (\ref{btfg}) associated to $F_g(\lambda)$, on the left, 
and of the series where we removed the constant map contribution $F_g(\lambda)-c_g$, 
on the right. When the constant map contribution is included, the leading singularity in the Borel plane takes place 
in the imaginary axis, near $\pm 2 \pi^2 \ri$, as expected from (\ref{cm-inst}). The subleading singularity, which we 
have indicated by a blue circle, corresponds precisely to the value of the instanton action (\ref{s-inst}), 
\be
A_s(\lambda\approx 2.61) \approx  44.73 + 9.87 \ri.
\ee
When the constant map contribution is removed, this becomes the leading singularity. 
%
%
Finally, the singularities 
display a periodicity of $2 \pi^2$ in the imaginary direction, corresponding to multiples of the constant period. 
Note as well that all singularities come in groups of four, $A$, $-A$, $A^*$ and $-A^*$. This is due to the fact that the perturbative series only has even powers (hence the 
parity symmetry $A \rightarrow -A$) and it is real (hence the conjugation symmetry $A\rightarrow A^*$). 
In overall, these numerical results are in good agreement with the analytic results conjectured in \cite{dmpnp}.

\begin{figure}
\begin{center}
\includegraphics[scale=0.5]{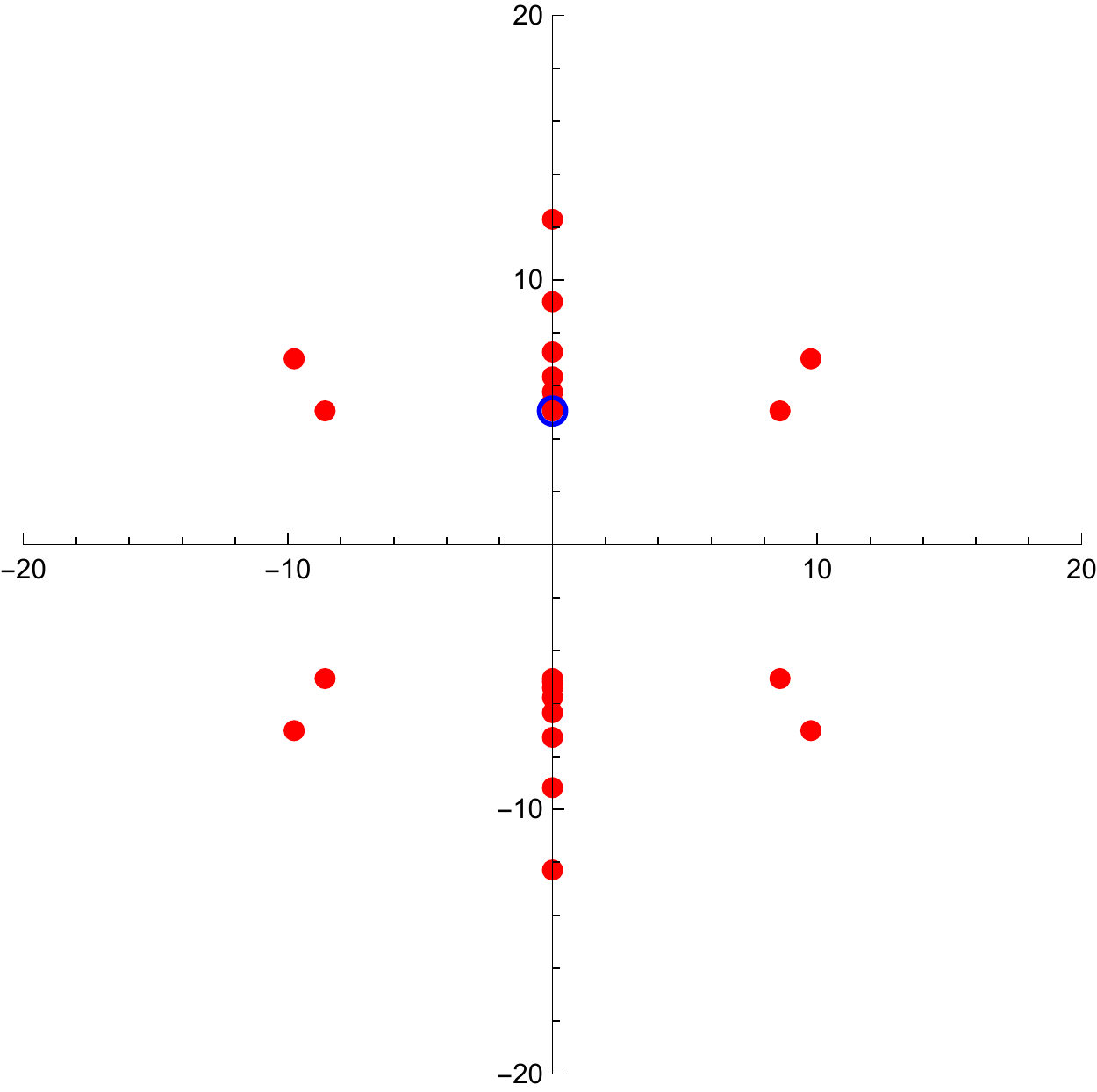} 
\caption{The red dots signal the location of the 
poles of the Pad\'e approximant (\ref{paden}), for the series $F_g(\lambda)-c_g$ and $\lambda\approx 0.128$. The blue circle corresponds to the numerical value of the purely imaginary instanton action 
$A_w(\lambda)$ in (\ref{w-inst}). The degree of the Pad\'e approximant is $n=54$.}
\label{wcpole}
\end{center}
\end{figure}

 The above analysis can be repeated for smaller values of $\lambda$. According to \cite{dmpnp}, if $\lambda$ is sufficiently small, 
 the closest pole to the origin should be 
 located at (\ref{w-inst}), and this is precisely what is found from the analysis of the Pad\'e approximants. As an example, we show in \figref{wcpole} 
 the poles of the Pad\'e approximant of the series $F_g(\lambda)-c_g$, for $\lambda=0.128$. The blue circle corresponds precisely to the value of $A_w(\lambda)$, and gives 
 the location of the leading singularity. We conclude that the structure of the poles in the Borel plane agrees with the analysis in \cite{dmpnp}: when $\lambda$ is small (weak coupling 
 regime), the leading singularity in the Borel plane for the sequence $F_g(\lambda)-c_g$ is given by the instanton action (\ref{w-inst}). As we increase the 't Hooft coupling and we enter 
 the strong coupling regime, the singularities move in the plane and the one corresponding to $A_s(\lambda)$ becomes dominant (i.e. smaller in absolute value).

One important aspect of the numerical structure of the poles of the Pad\'e approximants is that 
no singularities appear along the positive real axis\footnote{One might find poles in the positive real axis 
for some of the Pad\'e approximants, but they are not stable and they should be regarded as artifacts of the numerical approximation. When these accidental poles 
are present, we perform the Borel resummation by deforming the contour slightly above the real axis.}. This 
is again consistent with the analysis in \cite{dmpnp}. It indicates that the series is likely to be Borel summable, and therefore 
we can perform a standard Borel--Pad\'e resummation (\ref{analytic}): we do the integral (\ref{pade-approx}) 
for the Pad\'e approximant of (\ref{btfg}), and we add at the end the terms of genus zero and one. We will denote by 
\be
\label{fbpn}
F^{\rm BP}_{n} (N,k)
\ee
the final result, where the subindex $n$ refers to the degree of the Pad\'e approximant in (\ref{paden}).  

 In Table \ref{Tableorbifold} we present some of our numerical results, for various values of $N$ and $k$ (this also fixes the value of $\lambda$), and for $n=50$. 
 The first column shows the numerical value of $F^{\rm BP}_{n} (N,k)$. The second column shows the difference 
 between the exact results for the free energy listed in \cite{hmo2} and the Borel--Pad\'e resummation, i.e 
 \be
 \label{diffe}
 F(N,k)- F^{\rm BP}_{n} (N,k). 
 \ee
The third column gives an estimate of the error incurred in the Borel--Pad\'e resummation (since the values of $F(N,k)$ for the chosen $N$, $k$ are 
 known analytically, the error in their numerical value can be made arbitrarily small.)
 We see that the difference (\ref{diffe}) is systematically bigger than the error estimate. One is forced 
to conclude that {\it the Borel resummation of the $1/N$ expansion does not reproduce the expected exact value}. We are lacking 
non-perturbative information, and Borel summability is not enough to reconstruct the answer.

\begin{table} \begin{center}
 \begin{tabular}{|c|c|l|l|c|}\hline
	
	$k$ &  $N$ & $F^{BP}_{50}(N,k)$ & difference & error estimate  \\ \hline
	3 & 3 & -12.855\,641\,0  &1.7 $ \cdot 10^{-6} $ &  $ 10^{-8}$ \\ \hline
	3 & 4 & -19.875\,288\,9 & 3 $ \cdot 10^{-7} $ & $ 10^{-8}$ \\ \hline
	3 & 5 & -27.873\,671\,3 & $<$error  &  $ 10^{-8}$ \\ \hline
	3 & 6 & -36.745\,088\,489  &1.4 $ \cdot 10^{-8} $ & $ 10^{-10}$ \\ \hline
	3 & 7 & -46.411\,925\,432 &3.6 $ \cdot 10^{-9} $ &  $ 10^{-11}$ \\ \hline
	\\ \hline
	4 & 3 & -14.661\,864\,163 & -2.48 $ \cdot 10^{-7} $ &  $ 10^{-10}$ \\ \hline
	4 & 4 & -22.739\,393\,064\,7  & -3.63 $ \cdot 10^{-8} $ & $ 10^{-11}$ \\ \hline
	4 & 5 & -31.951\,567\,722\,7 & -6.3 $ \cdot 10^{-9} $ &  $ 10^{-11}$ \\ \hline 
	4 & 6 & -42.174\,874\,324\,1 & -1.2  $ \cdot 10^{-9} $ &  $ 10^{-11}$ \\ \hline 
	4 & 7 &  -53.318\,854\,310\,15  & -2.8  $ \cdot 10^{-10} $ &  $ 10^{-12}$ \\ \hline 
	\\ \hline
	6 & 3 & -17.465\,291\,856\,437 &5.35 $ \cdot 10^{-9} $ & $ 10^{-13}$ \\ \hline
	6 & 4 & -27.259\,498\,892\,850\,8 &4.860 $ \cdot 10^{-10} $ & $ 10^{-14}$ \\ \hline
	6 & 5 & -38.456\,656\,039\,963\,9 &5.48 $ \cdot 10^{-11} $ &  $ 10^{-14}$ \\ \hline 
	6 & 6 &-50.901\,544\,876\,270\,6 &7.3  $ \cdot 10^{-12} $ & $ 10^{-14}$ \\ \hline 
	6 & 7 & -64.481\,001\,864\,961\,13 &1.11  $ \cdot 10^{-12} $ & $ 10^{-15}$ \\ \hline 

     \end{tabular}
        \caption{The first column shows the numerical value of the resummed free energy (\ref{fbpn}) for $n=50$, $F^{BP}_{50}(N,k)$, 
        and for various values of $N$, $k$. The second column shows the difference (\ref{diffe}) between the exact free energy $F(N,k)$ and the Borel--Pad\'e resummation. 
        The last column gives an estimate of the error in the computation of the Borel-Pad\'e resummation.}
        \end{center}
        \label{Tableorbifold}
   \end{table}

What kind of non-perturbative information are we lacking? It is easy to see that, for fixed $\lambda$, the difference (\ref{diffe}) decreases exponentially with $k$, so 
it could be due to a non-perturbative effect in $1/k$ (which is essentially the string coupling constant). As we have pointed out in the previous section, and as we have seen in the example 
of the quartic oscillator, this effect might be due to complex instantons: the action (\ref{s-inst}) is complex and does not obstruct Borel summability, but its real part 
is positive and might lead to non-perturbative effects which should be added explicitly to the Borel--Pad\'e resummation. Indeed, it is easy to see qualitatively (i.e. at the level of orders 
of magnitude) that, for $n$ large enough, 
\be
\label{estim1}
 F(N,k)- F_n^{\rm BP}(N,k) \sim \cos\left( {{\rm Im}\left(A_s (\lambda)\right)  \over 2 \pi}k + \phi \right) \exp\left[ -{k \over 2 \pi} {\rm Re}\left(A_s (\lambda)\right)  \right], 
\ee
where $\phi$ is a phase. 

\begin{figure}\begin{center}
\includegraphics[scale=0.6]{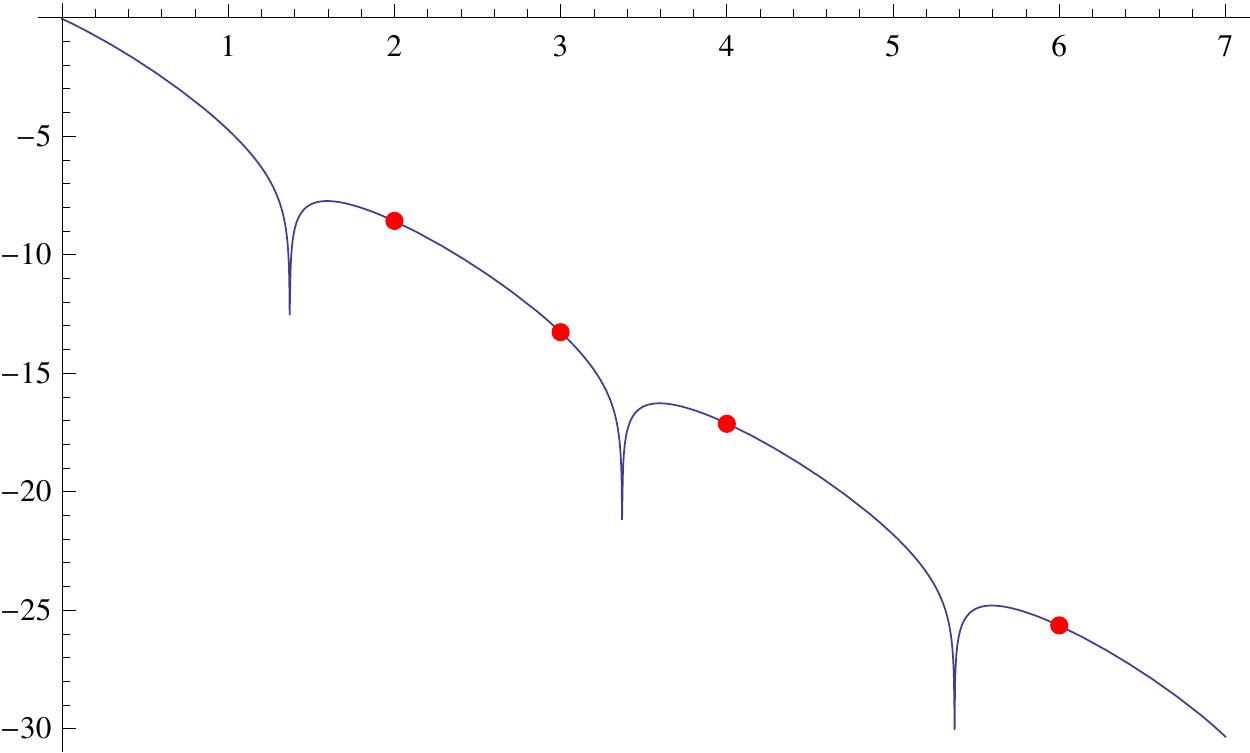}
\caption{This figures shows the four data points of $\log | F(N,k) - F_{46}^{\rm BP}(N,k)| $ against $k$, when $\lambda=1$, together with its optimal  fit to a function of the form (\ref{fitcos}), which 
turns out to be $0.13 - 4.27 k + \log\left| \cos\left(0.58- 1.57 k\right)\right|$. }
\label{fit}
\end{center}
\end{figure}

A more quantitative check of (\ref{estim1}) goes as follows: since we are working in the genus expansion, 
we should compare $F(N,k)$ and $F_n^{\rm BP}(N,k)$ at fixed 't Hooft parameter $\lambda$ but 
varying the string coupling constant $k$. Unfortunately, since the calculation of $F(N,k)$ is made for a limited 
range of integer values of $N$ and $k$, there are not that many data points with fixed $\lambda$. 
We have however four data points with $\lambda=1$ and $k=2,3,4,6$. Once $n$ is fixed, 
we can then fit the values of the l.h.s. of (\ref{estim1}), as we vary $k$, to a function of the form shown in the r.h.s. of 
(\ref{estim1}). This gives numerical estimates for the real and imaginary values of the action $A_s(\lambda)$, for a fixed degree of the 
Pad\'e approximant $n$. We then vary $n$ to extract a stable approximation, which can then be compared to the predicted values. In \figref{fit} we show the fit 
of 
\be
\log \left|  F(N,k)- F_{46}^{\rm BP}(N,k)  \right|
\ee
to a function of the form 
\be
\label{fitcos}
a + b k + \log \left|\cos\left(\phi + w k\right) \right|. 
\ee
Doing this for various values of the degree $n$ of the Pad\'e approximant, and keeping only the stable digits in the approximation, 
we get an estimate 
\be
A^{\rm fit}_s(\lambda=1) \approx 27 +  9.87 \ri, 
\ee
which should be compared to the expected value
\be
A_s (\lambda=1)\approx 27.33+ 9.87 \ri. 
\ee
This represents an error of $1$ percent, approximately, in the real part, 
and an even smaller one in the imaginary part of the action, which is quite good if 
we take into account that we only have four data points (also, in the fit we have assumed that the phase $\phi$ and proportionality coefficients in (\ref{estim1}) are 
constant, while they are expected to be slowly varying functions of $k$). Since ${\rm Im}(A_s(\lambda))= \pi^2$, the cosine function in (\ref{estim1}) is of the form 
\be
\label{cosk}
\cos\left( {\pi k \over 2} + \phi\right), 
\ee
and the sign alternation in the differences (\ref{diffe}), for fixed $\lambda$, is exactly as expected from (\ref{cosk}). 
 
We conclude that, although the string perturbation series in this example is Borel summable, it lacks crucial non-perturbative information to reproduce the exact answer. The general 
theory of non-perturbative effects tells us that we should consider a general trans-series incorporating the complex instantons with action (\ref{s-inst}). These instantons 
were interpreted in \cite{dmpnp} as due to D2 brane instantons (or membrane instantons in M-theory), so our conclusions are compatible with the analysis of the grand potential of ABJM theory in the 
M-theory expansion \cite{mp,hmo2,hmo3,cm,hmmo,km}, where it has been shown that membrane instantons are essential to make sense of the theory.

\sectiono{Resumming the genus expansion in topological string theory}

In this section we will consider a different, but related string perturbation series: 
the genus expansion of topological string theory on a particular local Calabi--Yau manifold, known as local $\IP^1 \times \IP^1$. 
This topological string theory has been studied in much detail, due to its relationship to Seiberg--Witten theory 
\cite{kkv}, to Chern--Simons theory on lens spaces \cite{akmv}, and to 
ABJM theory \cite{mpabjm, dmp}. In this section we will focus on the genus $g$ free energies $F_g$ 
in the so-called large radius frame. In this frame, the $F_g$s count holomorphic curves 
of genus $g$ in the Calabi--Yau target and they depend on two K\"ahler parameters, $T_1$ and $T_2$, 
which correspond to the (complexified) sizes of the $\IP^1$s. They have the structure 
\be
F_g(T_1, T_2) =  \sum_{d_1, d_2} N_{d_1, d_2}^g \re^{-d_1 T_1 - d_2 T_2}, 
\ee
where $N_{d_1, d_2}^g$ are the Gromov--Witten invariants of local $\IP^1 \times \IP^1$ at genus 
$g$ and for the degrees $d_1$, $d_2$. For $g\ge 2$ there is also a contribution due to constant maps, 
as in (\ref{cmap}), but this is much simpler to analyze and not relevant for our analysis. In the cases $g=0$ and $g=1$ there are also some additional contributions (which are cubic and linear polynomials in $T_{1,2}$, respectively), but these are also inessential to our purposes and will not be included in our definition of the $F_g$s.

 The free energies $F_g(T_1, T_2)$ can be computed in closed form by using the holomorphic anomaly equations of \cite{bcov}, adapted to the local case as in 
 for example \cite{hkr}. 
 However, their explicit calculation becomes difficult at higher genus. Therefore, we will consider them in the ``slice" of moduli space where 
\be
T_1=T_2=T. 
\ee
In that case, the functions $F_g(T)$ have the structure
\be
F_g(T)= \sum_{d\ge 1} N_d^g Q^d, 
\ee
where
\be
N_d^g= \sum_{d_1+d_2=d} N_{d_1, d_2}^g, \qquad  Q=\re^{-T}, 
\ee
and they can be computed in a much simpler way\footnote{The $F_g$s we will use have an additional factor of $4^{g-1}$ as compared to the standard ones 
in the topological string literature. This is equivalent to a rescaling $g_s \rightarrow 2 g_s$ of the string coupling constant.}. They are in fact modular transformations of the functions $F_g(\lambda)$ 
which we considered in the previous section 
\cite{dmp}. It will then be useful to use a parametrization related to the one we used there. Namely, we will use the 
mirror map 
\be
\label{m-map}
T=-4 z {~}_4F_3\left(1,1,\frac{3}{2},\frac{3}{2};2,2,2;16 z\right)-\log(z), 
\ee
where $z$ is related to the parameter $\kappa$ appearing in (\ref{lamkap}) by \cite{mpabjm,dmp}
\be
\label{z-kap}
z=-{1\over \kappa^2}. 
\ee
We will often parametrize the K\"ahler moduli space by 
\be
\label{qtau}
q=\re^{\ri \pi \tau_{\text{lr}}}, 
\ee
where 
\be \label{taulr}\tau_{\text{lr}}=\ri \frac{  K'\left(\frac{4 }{\ri \kappa }\right)}{K\left(\frac{4 }{\ri \kappa }\right)}.  
\ee
This large radius $\tau_{\rm lr}$ is related to the $\tau$ in (\ref{tauex}) through
\be \tau_{\text{lr}}-1=-{1\over \tau-1}. 
\ee
 The parametrization in terms of (\ref{qtau}) is more convenient since the $F_g$s are quasi-modular forms in the variable $\tau$. 
Of course, a given value of $q$ corresponds to a value of the K\"ahler parameter through the equations 
(\ref{m-map}), (\ref{z-kap}), (\ref{taulr}) and (\ref{qtau}). 

We would like to study the total topological string free energy 
\be
\label{ts-as}
F(T, g_s)= \sum_{g\ge 0} g_s^{2g-2} F_g(T),  
\ee
where $g_s$ is the topological string coupling constant. 
This is again an asymptotic series, for fixed $T$, and it behaves as 
\be
F_g(T) \sim \left| A_{\rm lr} (T)\right| ^{-2g} \cos\left( 2g\theta_{\rm lr} (T) +\delta_{\rm lr} (T) \right)(2g)!, 
\ee
where
\be
\label{alr}
A_{\rm lr}(T)= \pi T + 2 \pi^2 \ri, 
\ee
and we have written it as 
\be
A_{\rm lr}(T)=  \left| A_{\rm lr} (T)\right| \re^{\ri \theta_{\rm lr} (T) }.
\ee

It turns out that the total string free energy 
has a very different representation due to Gopakumar and Vafa \cite{gv2}. In 
this representation, one fixes the order of $\re^{-T}$ and resums the genus expansion. It has 
the structure
\be 
\label{st}
F^{\rm GV} (T,g_s )=\sum_{g \geq 0} \sum_{w,d \geq 1} n_g^d \left ( 2\sin \left( g_s w \right) \right ) ^{2g-2}\frac{1}{w} Q^{dw},
\ee
where $n_g^d$ are integers called Gopakumar--Vafa invariants. It is crucial to notice that the two series (\ref{ts-as}) and (\ref{st}) are very different. The first 
series should be understood as an asymptotic series in $g_s$ at fixed $T$. The series (\ref{st}) should be understood as a series in $Q=\re^{-T}$, with coefficients depending on $g_s$. 
Of course, when one expands the $F_g$s appearing in (\ref{ts-as}) in power series in $Q$, and when one expands the trigonometric functions in (\ref{st}) in powers of $g_s$, one obtains 
the same formal, double power series
\be
\sum_{g, d} N_d^g Q^d g_s^{2g-2}. 
\ee
One crucial question is then: what is the nature of the series appearing in the Gopakumar--Vafa representation? In \cite{hmmo}, 
numerical evidence was given that, surprisingly, if $g_s$ 
is {\it real}, the series (\ref{st}) has a finite radius of convergence in $Q$. 
However, the price to pay for this is the presence of an infinite number of poles in the real line: in fact, if we write 
\be
\label{gs-k}
g_s={2 \pi \over k}, 
\ee
then the series (\ref{st}) has double poles for any rational value of $k$. Since this is a dense set in $\IR$, the Gopakumar--Vafa representation does 
not seem to be very useful in the way of providing a non-perturbative definition of the 
theory, at least for real $g_s$. 

In the context of ABJM theory, one can relate the topological string free energy to the grand potential of the theory $J(\mu, k)$, which is defined by regarding $Z(N,k)$ as a
canonical partition function, i.e. 
\be
J(\mu,k) = \log \left( 1 + \sum_{N \ge 1} z^N Z(N,k) \right). 
\ee
As usual in Statistical Mechanics, the grand potential is a function of the chemical potential $\mu$, and we have also introduced the fugacity, 
\be
z=\re^\mu. 
\ee
The function $F^{\rm GV}(T,g_s)$ can be interpreted as the contribution from worldsheet 
instantons to the grand potential $J(\mu, k)$ \cite{hmo2}, where the relationship between $g_s$ and $k$ is given in (\ref{gs-k}), and 
\be
\label{T-mu}
T= {4 \mu \over k}-\ri \pi. 
\ee
In \cite{hmo2} it was pointed out that, since $Z(N,k)$ is well-defined for any value of $k$, 
there must be some additional contributions to $J(\mu, k)$ which cancel the divergences at rational $k$. 
These contributions are due to membrane instantons and they can be partially computed by using the Fermi gas approach of \cite{mp}. 
They were determined in the series of works \cite{hmo2,hmo3,cm,hmmo, km}. In particular, in \cite{hmmo} it was conjectured that they 
can be obtained from the refined topological string 
partition function \cite{ikv}, in the Nekrasov--Shatashvili limit \cite{ns}. In \cite{km} some aspects of this conjecture were derived from an analysis of the 
spectral problem appearing in the Fermi gas formulation. 

In order to set up the result for the non-perturbative completion of $F^{\rm GV}(T, g_s)$, we introduce the quantum-corrected K\"ahler parameter, 
\be
\label{teff}
T_{\rm eff}= T+ 2 \pi^2 \sum_{\ell \ge 1} a_\ell (k) \exp\left( -{k \ell \over 2}(T+ \ri \pi) \right). 
\ee
In this equation, the $a_\ell(k)$ are closely related to the coefficients of the quantum A-period of the local Calabi--Yau, which was introduced in \cite{mir-mor,acdkv} 
(see \cite{hkrs} for recent extensions). More details on these coefficients, as well as detailed values for the very first orders, can 
be found in for example \cite{hmmo}. We also introduce the membrane 
partition function, 
\be
\label{srt} 
F^{\rm M2}(T, g_s) = \sum_{\ell \geq 1}\left(a_\ell (k) \mu^2+b_\ell (k)\mu +c_\ell(k)\right)\re^{-2\ell\mu}.
\ee
Here, the relationship between $T, g_s$ and $\mu, k$ is the one expressed in (\ref{gs-k}) and (\ref{T-mu}). The coefficients $b_\ell (k)$ are closely related to 
the quantum B-periods of the local Calabi--Yau, and the coefficients $c_\ell(k)$ can be obtained from the coefficients $a_\ell(k)$, $b_\ell(k)$ \cite{hmo3,km}. 
We then define the non-perturbative topological string free energy as 
\be
\label{fnp}
F^{\rm NP}(T, g_s)= F^{\rm GV}(T_{\rm eff}, g_s) + F^{\rm M2}(T, g_s). 
\ee
The second term in (\ref{fnp}) is the membrane partition function. The first term, as compared to (\ref{st}), has additional contributions 
due to the promotion of $T$ to $T_{\rm eff}$. We will call these additional terms the contributions of bound states (of worldsheet instantons 
and membrane instantons). Note that the difference between (\ref{fnp}) and (\ref{st}) is purely non-perturbative in $g_s$, since the corrections to $T$ in (\ref{teff}), as well as the 
formal power series in $F^{\rm M2}(T, g_s)$, are of the form $\exp (-1/g_s)$. Therefore, the perturbative expansion of (\ref{fnp}) around $g_s=0$ agrees with (\ref{ts-as}). 
The non-perturbative free energy (\ref{fnp}) is in principle defined as a formal power series in the two small parameters 
$\re^{-T}$ and $\re^{-k T/2}$, with $g_s$ dependent coefficients. As we explained above, $F^{\rm GV}(T, g_s)$ has poles at all rational values of $k$. However, 
all these poles cancel in the function (\ref{fnp}), which as a formal power series is well-defined for all $g_s$ \cite{hmo2, hmo3, hmmo}. Moreover, 
from the analysis in \cite{hmmo} it seems that this series is convergent if $T$ is large enough. Therefore, at least for $T$ large, (\ref{fnp}) provides a 
well-defined non-perturbative completion of (\ref{st}). 

In the context of ABJM theory, the function (\ref{fnp}) has been argued to provide the exact 
grand potential $J(\mu, k)$ associated to the matrix model of ABJM theory. It contains non-perturbative effects which correct 
the perturbative answer given by $F^{\rm GV}(T, g_s)$. Although a complete derivation is still lacking, this proposal has passed 
many checks \cite{mp, hmo1, hmo2, cm, hmo3, hmmo, km}. The non-perturbative completion (\ref{fnp}) is motivated by its connection to 
the ABJM matrix model. 

We would like to understand the relationship between the asymptotic series (\ref{ts-as}) and the proposed non-perturbative answer (\ref{fnp}). More concretely, 
we would like to compare the Borel--Pad\'e resummation of (\ref{ts-as}), to the non-perturbative answer (\ref{fnp}). We will work in the slice of moduli space which corresponds 
to ABJM theory, namely $k$ and $\mu$ real. This in turn means that $g_s$ is real and that the imaginary part of $T$ is $-\ri \pi$. On this slice, the large order behavior of 
(\ref{ts-as}) is controlled by the complex instanton action (\ref{alr}). Since the leading singularity in the Borel plane is complex, we might have a Borel summable series.

\begin{figure}
\begin{center}
{\includegraphics[scale=0.6]{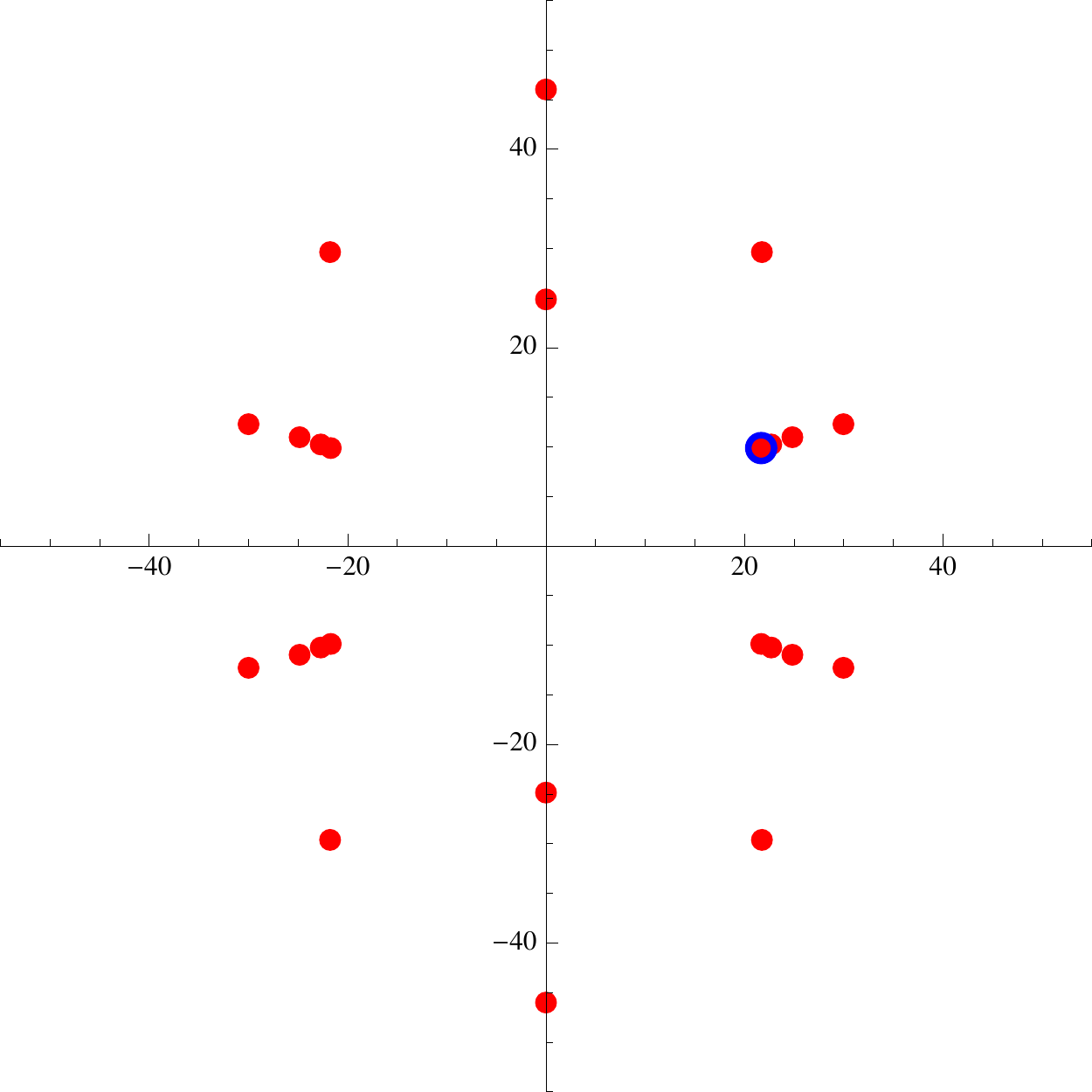}}
\caption{The red dots signal the location of the poles of the Pad\'e approximant (\ref{paden}) with $n=58$, for the series (\ref{bt-ts}) and 
a fixed value of $T$ specified by $q=-10^{-3}$ (where $q$ is defined in (\ref{qtau})). The blue circle indicates the numerical value of the 
complex instanton action $A_{\rm lr}$.}
\label{polesOfPade1000}
\end{center}
\end{figure}

We then proceed as in the previous section: we define the Borel--Pad\'e transform as 
\be
\label{bt-ts}
\widehat \varphi (\zeta)=\sum_{g\geq 2} \zeta ^{2g-2}{F_g(T)\over (2g-2)!},  
\ee 
and we consider its Pad\'e approximants. The first thing we can do is to examine their poles in the Borel plane. 
The results are shown in \figref{polesOfPade1000}, for the point in the moduli space with $q=-10^{-3}$ and for an approximant (\ref{paden}) with 
$n=58$. We notice that indeed, the pole which is closest to the origin agrees with the analytic value 
\be
\label{tac}
A_{\rm lr}  \approx 21.69 + 9.87 \ri. 
\ee
There are no stable poles on the positive real axis, so the series seems to be Borel summable, and we can perform a Borel--Pad\'e resummation of the series (\ref{ts-as}). Like 
before, we will denote by $F^{\rm BP}_n (T, g_s)$ the result of the resummation by using a Pad\'e approximant of order $n$. 

We can also compute the proposed non-perturbative answer (\ref{fnp}) for different values of $T$, $g_s$.  In this case, this answer is only known 
in the form of a conjecturally convergent series, therefore there is a numerical error associated to the truncation of this series. We estimate the error in $F^{\rm NP}(T, g_s)$ 
as follows: we first consider the series (\ref{fnp}), where the Gopakumar--Vafa series is truncated at order $Q^{10}$, and the series involving the non-perturbative effects 
is truncated at order $\ell=\left[{20 \over k} \right] $. Then we consider the sames series with truncation at $Q^{12}$ and $\ell=\left[{24 \over k} \right] $. The difference between these 
two results will be taken as a reliable error estimate\footnote{For $k=2$ we estimated the error by truncating the series at $Q^6$ and at $Q^4$}. 

\begin{table}
\centering
\begin{footnotesize}
 \begin{tabular}{|c|c|c|c|c|c|}\hline
      $k$ & $F^{\rm BP}_{54}(T, g_s)$  & error & $F^{\rm NP}(T, g_s) $ & error & difference\\ \hline
	2 & 0.0037 &  $ 10^{-5}$ &0.0056228650 &  $ 10^{-11}$ &0.0019\\ \hline %
	4 & 0.000996873 &  $ 10^{-10}$ &0.000993519297616245561182089 &  $ 10^{-28}$ &- 3.354 $\cdot 10^{-6}$ \\ \hline %
	6 & 0.0013366713318 &  $  10^{-14} $ & 0.0013366762433924625954836366 &  $  10^{-29} $ &4.9116  $ \cdot 10^{-9}$ \\ \hline %
	8 & 0.00200549863390460&  $  10^{-18} $ &0.0020054986273950134496944117&  $  10^{-29}$ & -6.50958  $ \cdot 10^{-12}$  \\ \hline
	10 &0.00290255648876704552 &  $  10^{-21} $ &0.002902556488775177021081745&  $  10^{-28}$ &8.13151  $ \cdot 10^{-15}$ \\ \hline
	12 &0.00401133227863213883147 &  $  10^{-24} $ &0.00401133227863212905949245&  $  10^{-27}$ &-9.77197  $ \cdot 10^{-18}$ \\ \hline
	14 &0.0053270579960529530912943& $  10^{-26} $ &0.00532705799605295310272304 &  $  10^{-27}$ &1.14287 $ \cdot 10^{-20}$ \\ \hline
	16 &0.0068479115274744906938481552&  $  10^{-29} $ &0.00684791152747449069383505 &  $  10^{-27}$ &-1.310 $ \cdot 10^{-23}$ \\ \hline
     \end{tabular}
     \end{footnotesize}
        \caption{In this table we show the values for the Borel-Pad\'e resummation of the series (\ref{ts-as}), for different integers $k$ and $q = -10^{-3}$, as well as the values of $F^{\rm NP}$ up to order $Q^{10}$. 
        In both cases we give an estimate of the numerical error. The last column shows the numerical value of the difference (\ref{estim}). }
           \label{LRcomp}
   \end{table}

\begin{figure}\begin{center}
{\includegraphics[scale=0.8]{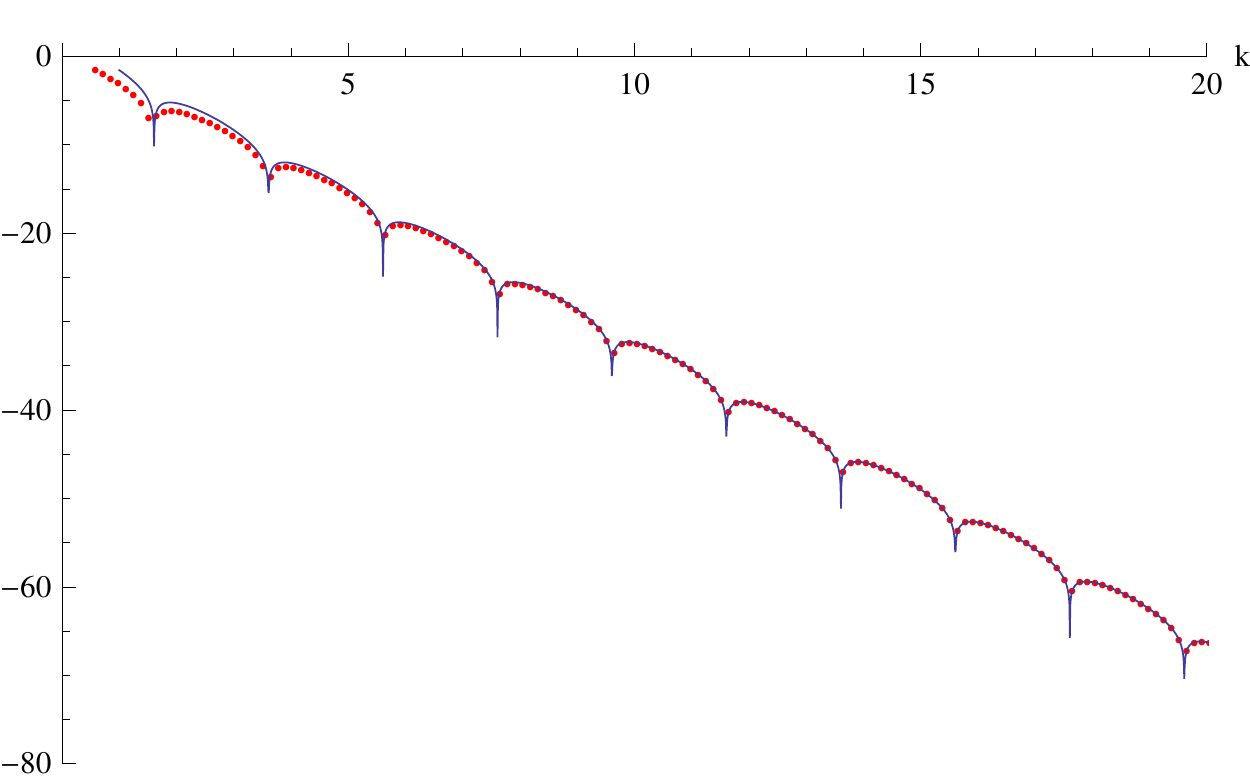} 
}
\caption{The red dots show the values of (\ref{logflr}), with $q=-10^{-3}$, and for 150 values of $k$. The blue line shows a fit of type (\ref{fitcos}) by using 75 points with $10 \leq k \leq 20$.
}
\label{Logdiff1000}
\end{center}
\end{figure}

We can now compare both results, the Borel--Pad\'e resummation and the non-perturbative result (\ref{fnp}). Our numerical results 
show conclusively that they are different. This is shown in table \ref{LRcomp}, where we consider a value of the K\"ahler parameter corresponding to 
$q = -10^{-3}$. As in the situation of the previous section, and in the case of the quartic oscillator, 
we interpret this difference as due to non-perturbative effects associated to complex instantons. The leading complex instanton has action given by (\ref{alr}), 
and we expect, for $n$ large enough, 
\be
\label{estim}
 F^{\rm NP} (T,g_s)- F_n^{\rm BP}(T,g_s) \sim \cos\left( {\rm Im}\left(A_{\rm lr} (T)\right){k  \over 2 \pi} + \phi \right) \exp\left[ -{k \over 2 \pi} {\rm Re}\left(A_{\rm lr} (T)\right)  \right]. 
\ee
where $\phi$ is a phase. In order to test this expectation, we can fix $T$, produce a sequence with the values of the l.h.s. of (\ref{estim}) as a function of $k$, and fit it to 
a function of the form shown in the r.h.s. In \figref{Logdiff1000} we show the fit 
of 
\be
\label{logflr}
\log \left|  F^{\rm NP} (T,g_s)- F_{54}^{\rm BP}(T,g_s)  \right|
\ee
to a function of the form (\ref{fitcos}). Doing this we get an estimate
\be
A^{\rm fit}_{\rm lr} \approx 21.27 +  9.87 \ri, 
\ee
which is to be compared to the expected value in (\ref{tac}). As we see, the difference is of about $2$ percent for the real part, 
and even smaller for the imaginary part. The error in the real part can be further reduced 
by using Richardson extrapolation. 
This procedure gives an estimate
\be 
{\rm Re }\left(A^{\rm fit}_{\rm lr} \right)\approx 21.61. 
\ee
Notice that the oscillation in the sign of the differences listed in Table \ref{LRcomp} is in precise agreement with the argument of the cosine, which is of the form 
\be
\cos\left( {\pi k \over 2} + \phi\right), 
\ee
since, for our choice of values of $T$, ${\rm Im}(A_{\rm lr}(T))= \pi^2$. 

We conclude that, even if the genus expansion of the topological string free energy is Borel summable, its Borel--Pad\'e resummation differs from 
the non-perturbative answer proposed in (\ref{fnp}) due to the presence of complex instantons, which should be incorporated directly, in the form of a trans-series. This is of course the same type 
of phenomenon which we observed in the case of the quartic oscillator and in the resummation of the $1/N$ expansion of ABJM theory.

There is an important point we would like to make concerning the Borel--Pad\'e resummation: although, when expanded in double power series of $g_s$ and $\re^{-T}$, the 
perturbative expansion (\ref{ts-as}) has the same information as the Gopakumar--Vafa representation, its resummation is perfectly smooth as a function of $g_s$. This 
seems to indicate that the HMO cancellation mechanism of \cite{hmo1} is already incorporated in 
the result of the resummation, and that the Gopakumar--Vafa representation 
of the topological string free energy (\ref{st}) has to be supplemented by some contribution which removes its poles. The contribution of 
membrane instantons and bound states in (\ref{fnp}) seems to be such that, after adding it to 
the contribution of worldsheet instantons, one obtains a quantity which is finite and equal to the Borel--Pad\'e 
resummation, up to exponentially small corrections given by the complex instantons. 

\begin{figure}
\begin{center}
{\includegraphics[scale=0.8]{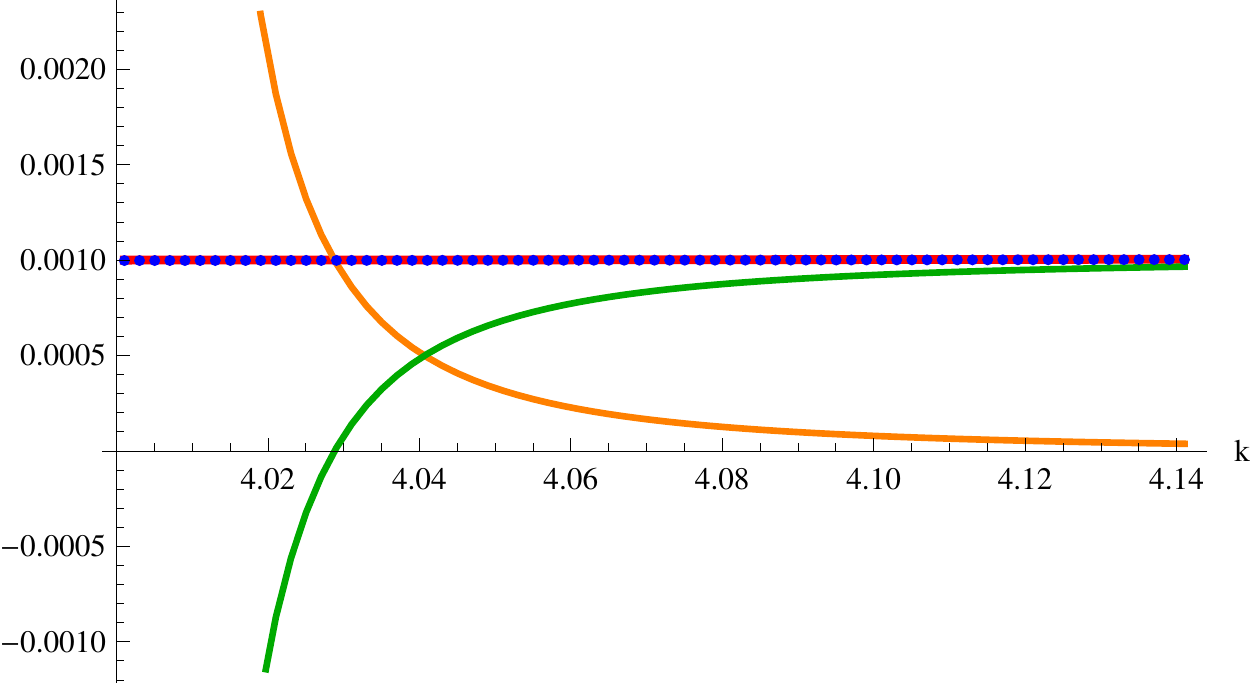}}
\caption{In this figure we show various relevant quantities for $q=-10^{-3}$, as a function of $k$, and near $k=4$. The orange line shows the non-perturbative contribution of membrane instantons and bound states. The green line shows the contribution of  worldsheet instantons, following from the Gopakumar--Vafa representations. The red line shows the Borel-Pad\'e resummation of the 
topological string series. The blue dots correspond to values of $F^{\rm NP}$, where we add worldsheet instantons and non-perturbative effects. The Gopakumar--Vafa series is truncated at $Q^{10}$ while the non perturbative series is truncated at  $\ell=\left[20 \over k\right]$. }
\label{pole-cancel}
\end{center}
\end{figure}

This can be seen in a very instructive way when we approach a pole at a small integer value of $k$, like for example $k=4$. 
In this case, the divergence in the Gopakumar--Vafa representation is seen already at order $Q^2$. Of course, the full Gopakumar--Vafa series diverges for any rational 
value of $k$, but we can truncate it to an appropriate order so that near $k=4$ only the divergence at this point is visible.
For instance by truncating the Gopakumar--Vafa serie at order $Q^{10}$, divergences occur at every $k$ of the form
\be
 k={w\over n}, \quad n \in \mathbb{N}, \quad  w\leq 10,
\ee
where  $w$ is the parameter appearing in (\ref{st}). In particular there is no divergence for $4 < k < 4.5$. At the same time, if $T$ is big enough, a truncation at order 
$Q^{10}$ leads to a very small error. Therefore, we can consider that the truncated Gopakumar--Vafa series gives a very good approximation to the exact answer, except that 
we have removed the non-perturbative contribution which regulates the pole at $k=4$. Indeed, this is what we observe 
in  \figref{pole-cancel}, where we show separately the contribution 
of worldsheet instantons, the non-perturbative contributions, the sum of both, and the Borel--Pad\'e resummation. First of all, we see that away from the pole, 
the non-perturbative corrections are very small; the worldsheet instanton contribution is the most important one and, 
up to exponentially small corrections, agrees with the Borel--Pad\'e resummation of the series. 
However, near the pole, the contribution of worldsheet instantons becomes very different from the value of the Borel--Pad\'e resummation. At the same time, 
the non-perturbative contributions become important and they have the right magnitude to give a total value of $F^{\rm NP}(T, g_s)$ close again to the Borel--Pad\'e resummation. 

\sectiono{Conclusions and prospects}

In this paper, by combining the AdS/CFT correspondence with results on the localization of ABJM theory and its $1/N$ expansion, 
we have been able to study the resummation of the string perturbation series in some examples, and compare it with the non-perturbative answer. 
Our main result is that, although the series seems to be Borel summable, it lacks explicit non-perturbative information due to the presence 
of complex instantons. This type of behavior appears in a much simpler model, first studied from this point of view by 
Balian, Parisi and Voros \cite{bpv}, namely the WKB series for the 
pure quartic oscillator in Quantum Mechanics. Although this WKB series is technically non-Borel summable, 
the leading, exponentially small error obtained in 
performing lateral Borel resummations is not 
due to the poles in the positive real axis, but to the poles associated to complex instantons. 

The results of this paper confirm that Borel summability is not a crucial property of asymptotic series. The key issue when faced 
with a perturbative scheme is whether we can 
extract the exact answer from {\it just} the perturbative series, or we have to include additional information. 
It is well-known that Borel summability is a necessary condition for this extraction, since when the series is not Borel summable 
one is forced to add non-perturbative sectors. However, being Borel summable is not a sufficient condition, since additional 
requirements are needed in order to reconstruct the original exact answer 
(like those appearing in Watson's theorem and its extensions). 
As we have argued in section 2, 
the mismatch between the Borel resummation of a Borel summable series and the exact answer is made possible by the presence 
of complex instantons. For this reason, this mismatch is not found in examples where complex instantons are absent, 
like the anharmonic quartic oscillator, where the Borel resummation agrees with the non-perturbative result \cite{ggs}.

It would be interesting to see in which cases the presence of complex instantons in a Borel-summable theory 
leads to a mismatch between Borel resummation and the non-perturbative result. Complex instantons seem to be necessary for 
this mismatch to occur, but they are not sufficient, and we know of an explicit example where this can be seen: 
in the $N$ vector model studied in \cite{hb}, we have verified that the 
Borel--Pad\'e resummation of the $1/N$ expansion of the free energy 
agrees with the exact result, in spite of the presence of complex instantons. A rich set of examples to study could come 
from non-unitary 2d CFTs coupled to 2d gravity. For example, the Yang--Lee singularity coupled to two-dimensional gravity leads to a Borel 
summable series for the specific heat, yet it contains complex instantons, and a non-perturbative definition 
has been proposed (see for example \cite{dfgzj} for a review and relevant references). To our knowledge, 
the Borel resummation of the series has not been compared in detail to the non-perturbative answer. Of course, 
beyond a list of examples and counter-examples, 
we would like to know if there is a simple criterium to determine in advance what is the relationship between 
the Borel resummation of the perturbative series and the exact answer. 

The next step in our research program would be to incorporate the complex instantons in an explicit way, 
through a trans-series ansatz. By the general theory of non-perturbative effects, 
we expect the exact answer to be given by the Borel resummation of a formal power series of the form (\ref{sigz}). 
This is given by the Borel resummation of the perturbative series, plus the Borel resummation of multi-instanton series, 
with certain weights which have to be determined. In the case of ABJM theory and topological strings, the most promising avenue 
for computing this formal trans-series is the formalism of \cite{cesv}, based on the holomorphic anomaly equations of \cite{bcov}, 
suitably extended to the non-perturbative sector. This would provide, in the context of the genus 
expansion, a detailed understanding of the non-perturbative effects in these theories. It would also make it possible to test some 
aspects of the proposal of \cite{hmmo} in models with no known large $N$ dual, like local $\IP^2$. 
We hope to report on these and related problems in the near future.

 \section*{Acknowledgements}
We would like to thank Ricardo Couso-Santamar\'\i a, Jean--Pierre Eckmann, Pavel Putrov, 
Ricardo Schiappa and Peter Wittwer for useful conversations. 
We are particularly grateful to Ricardo Schiappa for a detailed reading of the manuscript. 
This work is supported by the Fonds National Suisse, subsidies 200020-141329
and 200020-137523.

\end{document}